\documentclass[lettersize,journal]{IEEEtran}
\ifCLASSOPTIONcompsoc
  \usepackage[nocompress]{cite}
\else
  \usepackage{cite}
\fi
\ifCLASSINFOpdf
\usepackage[pdftex]{graphicx}
\graphicspath{.}
 \DeclareGraphicsExtensions{.pdf,.jpeg,.png}
\else
\fi

\usepackage{amsmath,amssymb,amsfonts, bbm, multirow, subfig}
\usepackage{color}
\usepackage{hyperref}

\usepackage[font=small]{caption}

\begin{document}
\title{Predicting Learning Interactions in Social \\ Learning Networks: A Deep Learning \\ Enabled Approach}

\author{Rajeev~Sahay${^*}$,~\IEEEmembership{Graduate Student Member,~IEEE,}
        Serena~Nicoll${^*}$,~\IEEEmembership{Student Member,~IEEE,}
        \\ Minjun Zhang, Tsung-Yen Yang,~\IEEEmembership{Student Member,~IEEE,} Carlee Joe-Wong,~\IEEEmembership{Member,~IEEE,} \\ Kerrie~A.~Douglas,~\IEEEmembership{Member, ~IEEE,} and Christopher~G.~Brinton,~\IEEEmembership{Senior~Member,~IEEE}
\IEEEcompsocitemizethanks{\IEEEcompsocthanksitem R. Sahay, S. Nicoll, M. Zhang, and C. G. Brinton are with the Elmore Family School of Electrical and Computer Engineering, Purdue University, West Lafayette, IN, 47907. E-mail: \{sahayr,snicoll,zhan3624,cgb\}@purdue.edu. \protect
\IEEEcompsocthanksitem K. A. Douglas is with the School of Engineering Education, Purdue University, West Lafayette, IN, 47907. \protect
E-mail: douglask@purdue.edu. 
\IEEEcompsocthanksitem T. Yang is with the Department of Electrical and Computer Engineering, Princeton University, Princeton, NJ 08544. E-mail: ty3@princeton.edu. \protect
\IEEEcompsocthanksitem C. Joe-Wong is with the Department of Electrical and Computer Engineering, Carnegie Mellon University, Pittsburgh, PA 15213. E-mail:cjoewong@andrew.cmu.edu. \protect
\IEEEcompsocthanksitem ${^*}$R. Sahay and S. Nicoll contributed equally to this work. \protect
\IEEEcompsocthanksitem This work was supported in part by the Charles Koch Foundation. \protect
\IEEEcompsocthanksitem The code and four of the datasets used in this work are available at \url{https://github.com/Jess-jpg-txt/sln-learning.} \protect
\IEEEcompsocthanksitem A preliminary version of the material in this work appeared in the Proceedings of the IEEE Conference on Computer Communications (INFOCOM) 2018 \cite{og}.}
}



\maketitle

\begin{abstract}
We consider the problem of predicting link formation in Social Learning Networks (SLN), a type of social network that forms when people learn from one another through structured interactions. While link prediction has been studied for general types of social networks, the evolution of SLNs over their lifetimes coupled with their dependence on which topics are being discussed presents new challenges for this type of network. To address these challenges, we develop a series of autonomous link prediction methodologies that utilize spatial and time-evolving network architectures to pass network state between space and time periods, and that models over three types of SLN features updated in each period: neighborhood-based (e.g., resource allocation), path-based (e.g., shortest path), and post-based (e.g., topic similarity). Through evaluation on six real-world datasets from Massive Open Online Course (MOOC) discussion forums and from Purdue University, we find that our method obtains substantial improvements over Bayesian models, linear classifiers, and graph neural networks, with AUCs typically above 0.91 and reaching 0.99 depending on the dataset. Our feature importance analysis shows that while neighborhood and path-based features contribute the most to the results, post-based features add additional information that may not always be relevant for link prediction.  
\end{abstract}




\section{Introduction}\label{sec:introduction}
\IEEEPARstart{O}{nline} education has exploded in popularity over the past few years, with estimates of up to 80\% of students having taken an online course \cite{slnsurvey}. The advent of the COVID-19 outbreak has significantly increased the number of online learners since 2020, which in turn has demonstrated online platforms' viability as an additional tool in physical classrooms. This growth has not been without challenges, however; online learning has raised concerns about its apparent lack of quality control, extraordinarily low teacher-to-student ratios, and scarcity of high-quality teachers \cite{slnsurvey}. The COVID-19 pandemic has highlighted the lack of quality tools for both students and teachers across online learning providers, making navigation of these massive communities a daunting or impossible task.

One way course providers have attempted to mitigate these problems is by establishing online forums where students can learn from each other, thus compensating for a lack of personalized instruction by posting questions, replying with answers, and otherwise exchanging ideas. Massive Open Online Courses (MOOCs), as well as Q\&A sites like Piazza, Quora, and StackOverflow, rely on forums extensively, generating a plethora of data about how users interact with one another online for learning purposes. These forums generate Social Learning Networks (SLNs) within communities of student users that evolve over time, facilitating peer-to-peer knowledge transfer in the absence of instructor intervention. Data-driven studies on the SLNs emerging from online learning forums have analyzed the benefits of social learning \cite{char_online_lrn,DiscussionForums} geared towards the ultimate goal of improving learning outcomes by, for example, proposing methods for instructor analytics \cite{moocperformance} and news feed personalization \cite{moocopt}.

In this work, we are motivated by the following research question: \emph{Can link formation between learners in an SLN be predicted in advance?} Such predictions would enable several new ways of improving online learning and forum experiences (e.g., encouraging early formation of learner groups or recommending that learners respond to newly-posted questions that they are expected to answer/contribute to later), thus helping to reduce the gap between in-person and online instruction. 

SLNs, however, pose two key challenges that differentiate them from standard time-evolving social networks \cite{dyn_graphs}. First, the SLN for an online course forms around the specific educational processes of that course \cite{sln_eff,edu_process}. With an SLN, users connect as a result of specific learning needs, and in response to events that are exogeneous to the discussion forum, e.g., the instructor releasing new content/assessments. On the other hand, homophily and pre-existing relationships are known to play a strong role in the evolution of standard social networks over time, which can provide initialization information for predicting learner interactions. An online SLN tied to a specific course, on the other hand, exhibits a ``cold start'' from a state of little-to-no observable network. Second, links in SLNs are defined much more arbitrarily compared to other graphs \cite{moocopt}. On social media sites, links between users are typically quantified with concrete metrics such as `friendships' or `follows,' where the connection between two users is explicit and typically optional. In an SLN, by contrast, a link between two users should indicate a transfer/sharing of knowledge. Explicit connection metrics do not typically exist, and even if they did, they do not imply the users have shared information. As a result of these challenges, the prediction of link formation in SLNs cannot be easily solved using previous methods designed for general time-evolving graphs \cite{graph_review}. 


In this work, we develop a link prediction methodology, specifically tailored for addressing the challenges associated with SLNs, which analyzes a set of features describing (i) learner pairs in an SLN and (ii) the evolution of learner interactions over time. Our methodology is deep learning-based, allowing consideration for both time-variable features and latent learner characteristics. We evaluate our methodology on data collected from four MOOC discussion forums from Coursera and two courses at Purdue University. We then investigate how our methodology can be used to make recommendations that may enhance the timing and quality of replies to discussion posts, thus encouraging interactions and improving learner experience in discussion-based forums.

\begin{figure*}[t]
\includegraphics[width = 18cm]{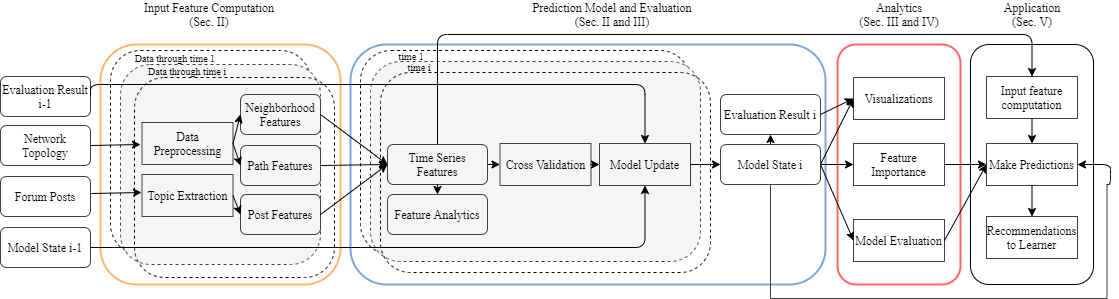}
\caption{Summary of the application of our SLN link prediction framework in post-based courses.}
\label{block_diagram}
\vspace{-0.15in}
\end{figure*}

\subsection{Related Work}\label{subsec:relatedwork}
The link prediction problem has been studied extensively in the context of online and digitally-enabled social networks, due to its usefulness in generating recommendations such as friendships, follows, or other forms of interactions \cite{sln_eff,m_slns,srn_eff,Survey}. Several methods have been proposed for this problem, beginning with unsupervised approaches and eventually transitioning to supervised methods in the past few years. In terms of unsupervised methods, \cite{unsup1} proposed using features based on node proximity and properties, while \cite{unsup2} and \cite{Jie2019} applied a model to incorporate additional contextual and temporal features. On the other hand, supervised approaches have proposed random walk algorithms using labels to increase the likelihood of traversing formed links \cite{randomwalk}, while \cite{supervised1} and \cite{supervised2} proposed deriving features from exogenous sources and training models on them to predict future link formation. Previous work has additionally considered using supervised and unsupervised methods simultaneously for exploratory learning environments \cite{sup_and_unsup}. However, these works do not consider characteristics unique to social \textit{learning} networks. Specifically, the potential dependence on discussion topics, and the need for time-series modeling is not explicitly modeled. Research into SLNs until this point has been largely theoretical, although \cite{casestudy} provides a first look into the application of deep learning-based link prediction algorithms in a classroom setting. Additionally, unsupervised approaches have demonstrated recent popularity for problems related classification of student behavior \cite{clustering}. Although the central focus of our research is concerned with SLNs, unlike these works, our strictly supervised models specifically consider student \textit{social} characteristics for large classrooms. 

Other works on online social networks have considered problems related to link formation, e.g., predicting the strength/repetition (rather than existence) of future links \cite{friendship,qanda1,Koprulu2019}, predicting link types \cite{clustering}, or examining the effects of student confusion on SLNs \cite{stu_conf}. The methods used and developed include linear regression/classification on network features and user demographics \cite{friendship,Zhang2020}, latent variable modeling of learner interaction frequencies \cite{clustering}, and dynamic models to account for the disappearance and strengthening of links over time \cite{supervised2}. Our models utilize some similar network features, but we consider the different prediction objective of pinpointing when links will form. In fact, given its high observed quality, we consider a time-series version of \cite{ogmodel} as a potential model.

An SLN is fully described by several datasets that each capture the a subset of student behavior inside the associated course. Recent papers choose to focus on one or a couple of these datasets: e.g. Student video-watching behavior \cite{moocperformance}, student performance \cite{Dalipi2018,Tsiamaki2020}, student physical behavior\cite{Yang2018}, or discussion forum data \cite{Papamitsiou2020,Almatrafi2019,Joksimovic2020,Marcos2019}. Our work is evaluated on a similar dataset to \cite{Joksimovic2020} in that it provides information gathered on student message passing behavior in a discussion forum. The models created in these other works fundamentally differ from our focus on individual student relationships. \cite{Papamitsiou2020} focuses on making group predictions from clusters of similar students, while \cite{Marcos2019} models changes in student behavior at critical points (e.g., exams and holidays).

Some recent works have focused on other aspects of different types of SLNs, e.g., MOOCs \cite{friendship,clustering}, \cite{MOOC}, Q\&A sites \cite{qanda1,qanda2}, and enterprise social networks \cite{socialnet1,socialnet2}. Our work is perhaps most similar to \cite{slnsurvey,friendship} in that we study prediction for SLNs using topological features. The prediction objectives in these other works, however, are fundamentally different than our focus of predicting interactions between learners in that they seek to predict course grades via video-watching behaviors \cite{MOOC} and student knowledge-state via learner post and reply frequencies \cite{qanda2}.

\subsection{Our Methodology and Contributions}
\label{subsec:methodcontributions}


In this work, we propose a novel framework specifically tailored to perform link prediction in SLNs. Fig. \ref{block_diagram} summarizes the main components of our methodology, which are further outlined in the following discussion. 

\subsubsection{Input Feature Computation}
We begin by extracting the discussion data from the considered forum to construct the SLN (Sec. \ref{sec:SLNgraphmodel}). Next, we engineer a set of features for each learner pair (Sec. \ref{sec:featureengineering}). Here, we define three groups of features that we consider: (i) neighborhood-based features that are determined from common neighborhoods, (ii) path-based features based on paths between learners, and (iii) post-based features that are determined from latent topic analysis of learner posts. Because a specific definition of what constitutes link formation between two users in an SLN does not exist, a key question when quantifying an SLN is how best to model learner interactions without loss of accuracy \cite{moocopt}. We address this through inference from forum data, with consideration for both quality of interaction \cite{ogmodel} and timing.

\subsubsection{Prediction Model}
The second component of our framework shown in Fig. \ref{block_diagram} is the prediction model (Sec. \ref{sec:linkpredictionmethodology}). We consider three different classes of predictors: (i) linear classifiers, (ii) graph neural networks (GNN), and (iii) gradient-based deep neural network classifiers (specifically, Bayesian neural networks, fully connected neural networks, convolutional neural networks, recurrent neural networks, and convolutional recurrent neural networks). The success of Bayesian models in static link prediction problems\cite{bayes} motivates us to consider their performance in the time-evolving SLN setting, while GNNs offer efficient learning over graphs without explicit feature engineering \cite{gnns}. However, we develop our core methodology around deep learning-based classifiers, because, as we will show, explicit feature modeling paired with various layer types, which can extract spatial or temporal patterns from the SLN features, result in more robust and accurate SLN link prediction. 

\subsubsection{Evaluation and Analytics}

To assess the quality of our models, we train and evaluate our considered prediction models on four MOOC discussion forums and two Piazza discussion forums, using an unsupervised method as a baseline (Sec. \ref{sec:baseline}). Through our evaluation, we also generate four types of analytics. The first analytic is feature importance, which quantifies the importance of each considered feature group. The second and third analytics quantify time-dependent model parameters, including closeness between time of link prediction and actual link formation as well as the relationship between features and the timing and quality of formed links. The fourth analytic explores the effects of varying classification architectures, where we anaylize the importance of different architectures in different course types (e.g., quantitative vs. humanities). In addition to these analytics, we provide visualizations for instructors to interact with the results of our proposed framework and respond to changes in the course SLN. These visualizations encapsulate our analytics, allowing for interpretation by those not familiar with our model.

\textbf{Summary of Contributions:} In summary, our contributions are (i) developing a link prediction framework for SLNs, which learns based on topological and post-based features of user discussions (Sec. \ref{sec:SLNmodel}), (ii) demonstrating that the combination of our features with spatial pattern-capturing neural networks obtains the most robust SLN link prediction quality over six datasets, with AUCs above $0.90$ in each case (Sec. \ref{sec:modelevaluation}), and (iii) developing a set of analytics for SLN link formation based on our link prediction framework (Sec. \ref{sec:recommending}).


\section{Social Learning Network Methodology}
\label{sec:SLNmodel}
In this section, we formalize our SLN link prediction methodology. We first quantify an SLN from forum data (Sec. \ref{sec:SLNgraphmodel}) and define the particular features that are used as model inputs (Sec. \ref{sec:featureengineering}). We then develop unsupervised predictor, linear classifiers, GNNs, and deep learning classifiers (Sec. \ref{sec:linkpredictionmethodology}) for link prediction.

\subsection{SLN Graph Model}
\label{sec:SLNgraphmodel}
In order to define our features, we must first describe how link creation in an SLN model is inferred and quantified from online forum data. 

\subsubsection{Online forums} \label{sec:olforums} 

The format of online forums differs by host site and by classroom needs. We identify two main types of forum structures to account for in our methodology. 

\textbf{MOOC forum structure:} A large online forum such as those hosted on Coursera is typically comprised of a series of threads, with each thread in turn being comprised of one or more posts. Each post is written by a single user. A post, in turn, can have one or more comments attached to it. Given the observation that SLN forum users do not abide by the designation of post vs. comment consistently \cite{moocopt}, we will not distinguish between them, instead referring to them both as posts. This structure of thread posts is depicted in Fig. 2a.

\textbf{Q\&A forum structure:} Another format, implemented by Piazza, forces a ``Question/Answer" thread structure. The forum is constructed from a series of questions and their responses, with allowance for follow-up questions and responses. In contrast to traditional forums, a response on Piazza may have contributions from multiple users in the same block, rather than requiring a new comment from each user. Any question may have comments attached to it in the form of ``follow-ups", which can in turn generate new responses. Using the observation listed above from \cite{moocopt} again, we do not distinguish between types of follow-up responses and label all responses after the initial question as posts. This alternate structure of thread posts is depicted in Fig. 2b.

\subsubsection{Quantifying SLN link creation}\label{sec:quantSLN} A link $(u, v)$ is observed between learner $u$ and another learner $v$ if, in a specific time interval, both $u$ and $v$ contribute to a post in the same thread (e.g., by either creating the initial post or contributing via a follow-up post). We use this as the criterion for establishing the link $(u, v)$ in the SLN because it signifies the fact that learner $u$ and learner $v$ have exchanged ideas and interacted in the same thread within a specific time interval.

\begin{figure}[t]
\includegraphics[width=8cm]{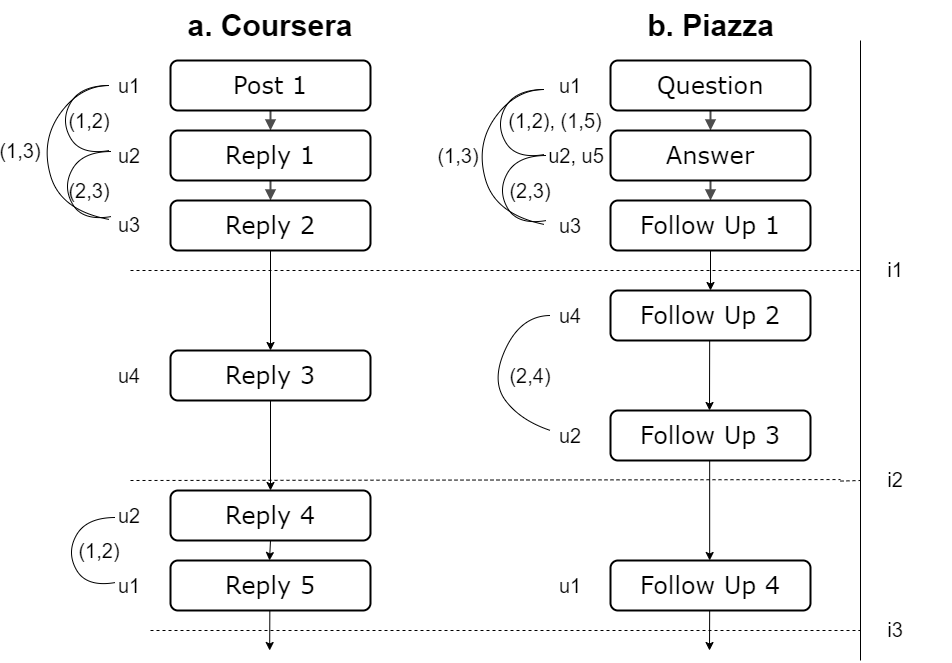} 
\caption{Example of how posts in two different forum structures are divided into time periods and how SLN link creation between the learners authoring these posts is modeled. Fig. 2a (left): model for a Coursera forum. Fig. 2b (right): model for a Piazza forum.} \label{posts}
\vspace{-0.15in}
\end{figure}

To model the evolution of an SLN, we group its posts into different time intervals. Specifically, we divide all posts in a given thread into $L$ equally spaced intervals. Fig. \ref{posts} illustrates this procedure for two example threads. We use $y_{uv}(i)$ as an indicator variable for the formation of link $(u, v)$: $y_{uv}(i) = 1$ if a link between $u$ and $v$ has been created in any interval up to and including $i$, and $y_{uv}(i) = 0$ otherwise. Thus, as in most social networks \cite{socialnet2}\cite{randomwalk}, links persist over time in our SLN model. The SLN graph structure in any given interval $i$ is then comprised of nodes corresponding to the learners $u$ and edges $(u, v)$ corresponding to links between them. For the purpose of predicting future responses, we consider this interaction to be bidirectional, i.e., the resulting SLN is an undirected graph. Formally, we define $\mathcal{G}(i) = [y_{uv}(i)]$ as the binary adjacency matrix of the SLN during interval $i$; since links are bidirectional, $\mathcal{G}(i)$ is symmetric. 

\begin{figure}[t]
    \centering
    \includegraphics[width=0.45\textwidth]{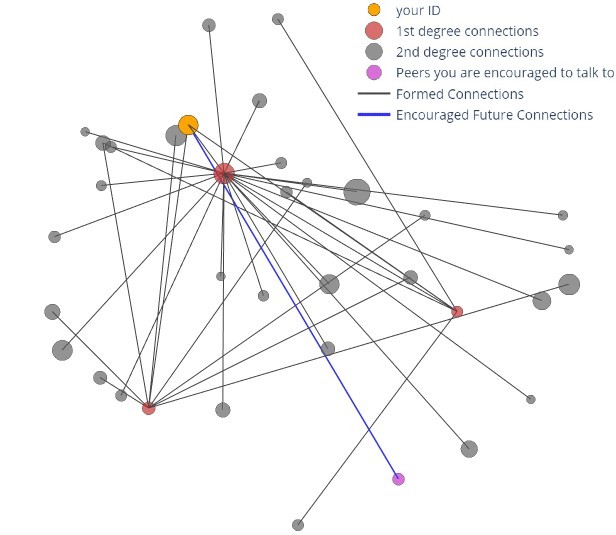}
    \caption{A snapshot of the SLN graph model for a single user (represented by a unique ID string) and their close neighborhood. The visual demonstrates the lack of multiple paths between users, underlying the sparse nature of the graph. }
    \label{user_view}
    \vspace{-0.15in}
\end{figure}

We can also define subgraphs of $\mathcal{G}(i)$ focusing on particular students. Fig. \ref{user_view} visualizes the neighborhood for an individual, randomly selected student at a particular time instance, where first and second degree connections are considered. In addition to capturing detailed link-formation behavior evaluated later in this study, evaluating a visual representation from the perspective of a single student provides an intuition for individual student contributions and demonstrates the presence of ``hub" students. The lack of multiple paths between students highlights the underlying sparse nature of $\mathcal{G}(i)$, requiring users to traverse one long path rather than choose from several short connections. Additionally, the relative small false positive rate (denoted by blue links in Fig. \ref{user_view}) demonstrates our framework's efficacy for link prediction, as we will describe further in Sec. \ref{sec:LPE}.

Two particular subsets of $\mathcal{G}(i)$ are of interest in the link prediction problem. We define
\begin{equation}
    \Omega = {(u, v) : u, v \in N(\mathcal{G}), u \neq v},
\end{equation}
i.e., all possible learner pairs in the SLN. We then define two subsets of $\Omega: \mathcal{G}(L)$, which is the set of formed links at the final time $i = L$ (i.e., with $y_{uv}(L) = 1$), and $\mathcal{G}^{c} (L) = \Omega \hspace{1mm} \backslash \hspace{1mm} \mathcal{G}(L)$, the complement graph of un-formed links (i.e., $y_{uv}(L) = 0$). Note that $|\mathcal{G}^{c}(L)|\gg|\mathcal{G}(L)|$ for each dataset (i.e., most learners are never linked). This large class imbalance between formed and unformed links informs our link prediction framework in Sec. \ref{sec:linkpredictionmethodology}. 


\subsection{SLN Feature Engineering}
\label{sec:featureengineering}

\begin{table*}[t]
\centering
\begin{tabular}[width = 16cm]{c c c c c c c c} 
 \hline
 Forum & Course Title & Beginning & Duration & Users & Threads & Learner Pairs & Posts \\
  \hline
 \hline
 $\mathtt{ml}$ & Machine Learning & 4/29/13 & 12 & 4263 & 4217 & 73315 & 25481 \\ 
 \hline
  $\mathtt{algo}$ & Algorithms: Design and Analysis I & 9/22/14 & 13 & 3013 & 4656 & 50006 & 16276 \\
  \hline
 $\mathtt{shake}$ & Shakespeare in Community & 4/22/15 & 5 & 958 & 1389 & 66217 & 7484 \\ 
 \hline
  $\mathtt{comp}$ & English Composition I & 7/01/13 & 8 & 1862 & 1286 & 20083 & 8255  \\
  \hline
   $\mathtt{f19}$ & Python for Data Science & 8/20/19 & 18 & 115 & 669 & 17000 & 2013 \\
   \hline
   $\mathtt{s20}$ & Python for Data Science & 1/17/20 & 17 & 290 & 1129 & 44964 & 4955\\
 \hline
\end{tabular}
\caption{Descriptive metrics on our six considered forum datasets. The title, beginning date (m/dd/yy), duration (weeks), number of users, threads, learner pairs, and posts by the end. All courses were broken into 20 time instances.}
\label{course_stats}
\vspace{-0.15in}
\end{table*}

We now define our features, computed for each learner pair $(u, v), u \neq v$. These quantities serve as the inputs to our prediction algorithms in Sec. \ref{sec:linkpredictionmethodology}. 

\textbf{Neighborhood-based Features}: These features, as well as path-based features discussed next, are extracted from the topology of the graph. Letting $N(\mathcal{G})$ be the set of nodes in the SLN $\mathcal{G}$ and $\Gamma_{u}(i) \subseteq N(\mathcal{G})$ denote the set of neighbors of $u$ at time $i$, the neighborhood-based features qualitatively measure the ``similarity” of $u$ and $v$’s neighborhoods \cite{NPLP}. They are quantified as follows:

\begin{enumerate}
    \item \emph{Jaccard coefficient}:
    $$\texttt{Ja}_{uv} = |\Gamma_{u}(i) \cap \Gamma_{v}(i)| / |\Gamma_{u}(i) \cup \Gamma_{v}(i)|$$
    \item \emph{Adamic-Adar index}:
    $$\texttt{Ad}_{uv} = \sum_{n \in \Gamma_{u}(i) \cap \Gamma_{v}(i)} 1 / \text{log} |\Gamma_{n}(i)|$$
    \item \emph{Resource allocation index}:
    $$\texttt{Re}_{uv} = \sum_{n \in \Gamma_{u}(i) \cap \Gamma_{v}(i)} 1 / |\Gamma_{n}(i)|$$
    \item \emph{Preferential attachment score}: $$\texttt{Pr}_{uv} = |\Gamma_{u}(i)| \cdot |\Gamma_{v}(i)|$$
\end{enumerate}

We let $\mathbf{b}_{uv}$ denote the vector of these features for pair $(u, v)$. Note that a larger value of each of these features, roughly speaking, indicates that $u$ and $v$ share more common, low degree neighbors than they do with others.

\textbf{Path-based Features}: These features measure the proximity of $u$ and $v$ in the SLN. They are as follows:

\begin{enumerate}
    \setcounter{enumi}{4}
    \item \emph{Shortest path length} ($\texttt{Lp}_{uv}$): The length of the shortest path between $u$ and $v$.
    \item \emph{Number of paths} ($\texttt{Np}_{uv}$): The number of shortest paths (i.e., of length \texttt{Lp}) between $u$ and $v$.
\end{enumerate}

We let $\mathbf{a}_{uv}$ denote the vector of these features. Note that as \texttt{Lp} decreases, $u$ and $v$ become more closely connected, while a larger \texttt{Np} indicates more redundancy in these paths.

\textbf{Post-based Features}: Besides topology-based attributes, learners' interests in different course topics will also influence their probability of forming links in an SLN. In particular, we would expect those with similar topic interests to be more likely to post in the same thread, i.e., form links. We thus compare the topics of different learners' posts to compute another feature that shows the learners' similarity in interests.

To do this, we apply the Latent Dirichlet Allocation (LDA) algorithm \cite{LDA} on the dictionary of all course words (i.e., all unique words used in all the considered posts of a course) to extract a set, $\mathcal{K}$, of latent topics across posts, and a model of posts as a probability vector of these topics. In our application, we view each post as a separate ``document,” since learners are likely to discuss many distinct topics over time. For each learner, $u$, we obtain the latent topic vector of their posts through time $i$ as the average of their post vectors through $i$. We denote the set of topics for learner $u$ that exceed a minimum threshold of coverage across their posts through time $i$ as $K_{u}(i)$. With this, we define the last feature which captures the number of common topics between learners $u$ and $v$:
\begin{enumerate}
    \setcounter{enumi}{6}
    \item \emph{Number of common topics} (\texttt{To}): $|K_{u}(i) \cap K_{v}(i)|$
\end{enumerate}

We use $c_{uv}$ as the time-series version of \texttt{To}, i.e., the number of common topics discussed by $u$ and $v$. 

    

\subsection{Link Prediction Methodology}
\label{sec:linkpredictionmethodology}
As discussed in Sec. \ref{sec:featureengineering}, the features extracted from the graph topology contain spatially and temporally correlated patterns between learner pairs. Therefore, we employ prediction models that are capable of exploiting these patterns for accurate link prediction. In this capacity, we consider the efficacy of four distinct deep learning architectures for our proposed framework: (i) the fully connected neural network (FCNN), which offers effective latent space prediction; (ii) the convolutional neural network (CNN), which is highly effective for processing spatially correlated patterns; (iii) the long-short-term memory (LSTM) based recurrent neural network (RNN), which is desirable for time-series modeling; (iv) the convolutional recurrent neural network (CRNN), which extracts both spatial and temporal correlations. As baselines to these methods, and to demonstrate the necessity of the aforementioned classifiers and their corresponding architectures, we compare our proposed deep learning prediction framework to five traditional prediction models: an unsupervised predictor, two linear prediction models (support vector machines and linear discriminant analysis), a graph neural network \cite{lp_gnn}, and a Bayesian neural network \cite{bayes}.

For a given pair of users $(u, v)$, the input feature vector into each of the following models is given by $\mathbf{e}_{uv} = [\mathbf{b}_{uv}, \mathbf{a}_{uv}, c_{uv}]$  while the target output is the link state $y_{uv}(i) \in \{0, 1\}$. In the following, we describe the latent state of each model as well as their corresponding training procedures.

\subsubsection{Unsupervised Predictor}
\label{sec:baseline}

We begin by using a simple prediction algorithm as a benchmark for the parameter-based models described below. Choosing the feature most associated with link formation, we follow \cite{randomwalk} and turn the resource allocation index (\texttt{Re}) feature into an unsupervised predictor. To do this, we compute \texttt{Re} for each $(u, v) \in \Omega$, normalize the vector of values to $[0,1]$, and use this as $\hat{y}_{uv}(i)$.

\subsubsection{Linear Classifiers} 
\label{sec:linear} 

Next, we consider two relatively simple linear models for SLN link prediction: linear discriminant analysis (LinDA) and support vector machines (SVMs). Both models attempt to find a separating linear hyper-plane between learners who did and did not form links. However, both models are learned using different methodologies. Specifically, LinDA uses every sample during training and assumes samples in each class follow the same distribution and have the same covariance matrix whereas SVM makes no prior assumptions on the data's distribution and aims to find a decision boundary using the points that result in the highest error.

\subsubsection{Graph Neural Networks (GNN)} GNNs are a class of neural networks for learning over datasets expressed as graphs. They have been employed to perform link prediction on a variety of graph topologies \cite{lp_gnn,gnns}. A potential advantage of GNNs in our setting would be obviating much of the feature engineering in Sec. II-B, as they can learn directly from the graph structure. Thus, we compare the efficacy of GNNs to our proposed method for predicting link formation in SLNs. Specifically, we adopt a two-layer convolutional GraphSAGE model \cite{graphsage}, where node attributes of the SLN are self-generated during training. Here, the adjacency matrix of the SLN is used as input into the GraphSAGE model at a given time in order to predict future links.

\begin{table*}[t]
\centering
\subfloat[$\mathtt{ml}$]{%
\begin{tabular}{c|c r r} 
 Features & SNR & Mean & s.d\\
 \hline
 \multirow{2}{*}{$\mathtt{Ja}$} & \multirow{2}{*}{0.5741} & 0.1467 & 0.1818 \\ 
  & & 0.0224 & 0.0345 \\
  \hline
  \multirow{2}{*}{$\mathtt{Ad}$} & \multirow{2}{*}{0.8069} & 2.6963 & 2.6556 \\ 
  & & 0.2121 & 0.4783 \\
  \hline
  \multirow{2}{*}{$\mathtt{Re}$} & \multirow{2}{*}{0.8221} & 0.2838 & 0.3108 \\ 
  & & 0.0085 & 0.0241 \\
  \hline
  \multirow{2}{*}{$\mathtt{Pr}$} & \multirow{2}{*}{0.3478} & 5413.9 & 12436 \\ 
  & & 512.37 & 1653.8 \\
  \hline
  \multirow{2}{*}{$\mathtt{Lp}$} & \multirow{2}{*}{-0.7037} & 0.8712 & 0.3454 \\ 
  & & 1.6186 & 0.7165 \\
  \hline
  \multirow{2}{*}{$\mathtt{Np}$} & \multirow{2}{*}{-0.1603} & 2.0779 & 9.1893 \\ 
  & & 9.3004 & 35.855 \\
  \hline
  \multirow{2}{*}{$\mathtt{To}$} & \multirow{2}{*}{0.2019} & 1.0201 & 1.6955 \\ 
  & & 0.4904 & 0.9276 \\
 \hline
\end{tabular}}
\quad
\subfloat[$\mathtt{algo}$]{%
\begin{tabular}{c|c r r} 
 Features & SNR & Mean & s.d\\
 \hline
 \multirow{2}{*}{$\mathtt{Ja}$} & \multirow{2}{*}{0.6614} & 0.2312 & 0.2727 \\ 
  & & 0.0246 & 0.0396 \\
  \hline
  \multirow{2}{*}{$\mathtt{Ad}$} & \multirow{2}{*}{0.8254} & 3.1919 & 3.3436 \\ 
  & & 0.1748 & 0.3116 \\
  \hline
  \multirow{2}{*}{$\mathtt{Re}$} & \multirow{2}{*}{0.9411} & 0.3503 & 0.3355 \\ 
  & & 0.0092 & 0.0268 \\
  \hline
  \multirow{2}{*}{$\mathtt{Pr}$} & \multirow{2}{*}{0.3812} & 1797.6 & 3253.4 \\ 
  & & 270.87 & 752.06 \\
  \hline
  \multirow{2}{*}{$\mathtt{Lp}$} & \multirow{2}{*}{-0.6638} & 0.7974 & 0.3091 \\ 
  & & 1.4348 & 0.6511 \\
  \hline
  \multirow{2}{*}{$\mathtt{Np}$} & \multirow{2}{*}{-0.2191} & 1.3389 & 3.8776 \\ 
  & & 4.9092 & 12.421 \\
  \hline
  \multirow{2}{*}{$\mathtt{To}$} & \multirow{2}{*}{0.1668} & 0.5875 & 0.9624 \\ 
  & & 0.3364 & 0.5426 \\
 \hline
\end{tabular}}
\quad
\subfloat[$\mathtt{comp}$]{%
\begin{tabular}{c|c r r} 
 Features & SNR & Mean & s.d\\
 \hline
 \multirow{2}{*}{$\mathtt{Ja}$} & \multirow{2}{*}{0.2535} & 0.1608 & 0.2207 \\ 
  & & 0.0721 & 0.1291 \\
  \hline
  \multirow{2}{*}{$\mathtt{Ad}$} & \multirow{2}{*}{0.7276} & 1.8286 & 2.1686 \\ 
  & & 0.0956 & 0.2131 \\
  \hline
  \multirow{2}{*}{$\mathtt{Re}$} & \multirow{2}{*}{0.7648} & 0.2959 & 0.3434 \\ 
  & & 0.0045 & 0.0376 \\
  \hline
  \multirow{2}{*}{$\mathtt{Pr}$} & \multirow{2}{*}{0.3836} & 1041.8 & 2325.5 \\ 
  & & 38.123 & 291.32 \\
  \hline
  \multirow{2}{*}{$\mathtt{Lp}$} & \multirow{2}{*}{-0.7048} & 0.9248 & 0.3497 \\ 
  & & 1.8233 & 0.9251 \\
  \hline
  \multirow{2}{*}{$\mathtt{Np}$} & \multirow{2}{*}{-0.2498} & 1.3182 & 3.2174 \\ 
  & & 5.9579 & 15.352 \\
  \hline
  \multirow{2}{*}{$\mathtt{To}$} & \multirow{2}{*}{0.1258} & 0.5703 & 0.8587 \\ 
  & & 0.4039 & 0.4637 \\
 \hline
\end{tabular}}

\subfloat[$\mathtt{shake}$]{%
\begin{tabular}{c|c r r} 
 Features & SNR & Mean & s.d\\
 \hline
 \multirow{2}{*}{$\mathtt{Ja}$} & \multirow{2}{*}{0.3527} & 0.1354 & 0.1318 \\ 
  & & 0.0565 & 0.0914 \\
  \hline
  \multirow{2}{*}{$\mathtt{Ad}$} & \multirow{2}{*}{0.7148} & 2.6913 & 2.8538 \\ 
  & & 0.2612 & 0.5453 \\
  \hline
  \multirow{2}{*}{$\mathtt{Re}$} & \multirow{2}{*}{0.6648} & 0.2934 & 0.3647 \\ 
  & & 0.0143 & 0.0551 \\
  \hline
  \multirow{2}{*}{$\mathtt{Pr}$} & \multirow{2}{*}{0.4871} & 1904.1 & 3074.1 \\ 
  & & 142.67 & 541.58 \\
  \hline
  \multirow{2}{*}{$\mathtt{Lp}$} & \multirow{2}{*}{-0.7802} & 0.9519 & 0.2995 \\ 
  & & 1.7221 & 0.6874 \\
  \hline
  \multirow{2}{*}{$\mathtt{Np}$} & \multirow{2}{*}{-0.2414} & 1.8512 & 4.3331 \\ 
  & & 7.3385 & 18.397 \\
  \hline
  \multirow{2}{*}{$\mathtt{To}$} & \multirow{2}{*}{0.3151} & 1.3249 & 1.6287 \\ 
  & & 0.5906 & 0.7009 \\
 \hline
\end{tabular}}
\quad
\subfloat[$\mathtt{f19}$]{%
\begin{tabular}{c|c r r} 
 Features & SNR & Mean & s.d\\
 \hline
 \multirow{2}{*}{$\mathtt{Ja}$} & \multirow{2}{*}{0.5807} & 0.1413 & 0.1294 \\ 
  & & 0.0323 & 0.0582 \\
  \hline
  \multirow{2}{*}{$\mathtt{Ad}$} & \multirow{2}{*}{0.6414} & 1.8429 & 2.0376 \\ 
  & & 0.2099 & 0.5084 \\
  \hline
  \multirow{2}{*}{$\mathtt{Re}$} & \multirow{2}{*}{0.5999} & 0.2633 & 0.3315 \\ 
  & & 0.0241 & 0.0673 \\
  \hline
  \multirow{2}{*}{$\mathtt{Pr}$} & \multirow{2}{*}{0.6066} & 360.11 & 449.03 \\ 
  & & 32.847 & 90.413 \\
  \hline
  \multirow{2}{*}{$\mathtt{Lp}$} & \multirow{2}{*}{-1.1082} & 1.3231 & 0.3538 \\ 
  & & 2.1759 & 0.4158 \\
  \hline
  \multirow{2}{*}{$\mathtt{Np}$} & \multirow{2}{*}{-0.4079} & 1.7306 & 1.4857 \\ 
  & & 3.9584 & 3.9746 \\
  \hline
  \multirow{2}{*}{$\mathtt{To}$} & \multirow{2}{*}{0.6042} & 2.6515 & 2.8861 \\ 
  & & 0.3702 & 0.8893 \\
 \hline
\end{tabular}}
\quad
\subfloat[$\mathtt{s20}$]{%
\begin{tabular}{c|c r r} 
 Features & SNR & Mean & s.d\\
 \hline
 \multirow{2}{*}{$\mathtt{Ja}$} & \multirow{2}{*}{0.6901} & 0.1341 & 0.1088 \\ 
  & & 0.0266 & 0.0468 \\
  \hline
  \multirow{2}{*}{$\mathtt{Ad}$} & \multirow{2}{*}{0.6628} & 2.5344 & 2.8694 \\ 
  & & 0.2289 & 0.6088 \\
  \hline
  \multirow{2}{*}{$\mathtt{Re}$} & \multirow{2}{*}{0.6149} & 0.2347 & 0.3019 \\ 
  & & 0.0164 & 0.0531 \\
  \hline
  \multirow{2}{*}{$\mathtt{Pr}$} & \multirow{2}{*}{0.5902} & 1109.7 & 1469.8 \\ 
  & & 81.076 & 273.16 \\
  \hline
  \multirow{2}{*}{$\mathtt{Lp}$} & \multirow{2}{*}{-0.9782} & 1.4292 & 0.3748 \\ 
  & & 2.1761 & 0.3887 \\
  \hline
  \multirow{2}{*}{$\mathtt{Np}$} & \multirow{2}{*}{-0.2908} & 2.9899 & 2.9203 \\ 
  & & 6.0636 & 7.6483 \\
  \hline
  \multirow{2}{*}{$\mathtt{To}$} & \multirow{2}{*}{0.6691} & 2.8634 & 2.9075 \\ 
  & & 0.3679 & 0.8221 \\
 \hline
\end{tabular}}
\caption{Summary statistics – SNR, mean and standard deviation (s.d.) – for the network features of the two link groups. The top row for each feature corresponds to formed links ($y_{uv}(L) = 1$), and the bottom to non-formed links ($y_{uv}(L) = 0$). Taken individually, the neighborhood-based features Re and Ad have the strongest correlations with link formation, while the topic-based To tends to have the least.}
\label{summary_stats}
\vspace{-0.15in}
\end{table*}

\begin{table}
\centering
\scalebox{0.8}{%
\subfloat[$\mathtt{ml}$]{%
\begin{tabular}{c|r r}
k & Support & Top 3 Words \\
\hline
1 & 0.1257 & class question svm \\
\hline
2 & 0.1078 & computer work image \\
\hline
3 & 0.0895 & gradient set lambda \\
\hline
4 & 0.0835 & code problem exercise \\
\hline
5 & 0.0741 & octave line column \\
\hline
\end{tabular}}}%
\enspace
\scalebox{0.8}{
\subfloat[$\mathtt{algo}$]{%
\begin{tabular}{c|r r}
k & Support & Top 3 Words \\
\hline
1 & 0.2287 & thought fast graphs  \\
\hline
2 & 0.0872 & heap length max \\
\hline
3 & 0.0713 & algorithm time run \\
\hline
4 & 0.0684 & file sort merge  \\
\hline
5 & 0.0676 & set problem line \\
\hline
\end{tabular}}}%

\scalebox{0.8}{
\subfloat[$\mathtt{comp}$]{%
\begin{tabular}{c|r r}
k & Support & Top 3 Words \\
\hline
1 & 0.1141 & project composition https \\
\hline
2 & 0.0736 & annotated idea good \\
\hline
3 & 0.0541 & great word read \\
\hline
4 & 0.0486 & writ time read \\
\hline
5 & 0.0425 & feedback hope find  \\
\hline
\end{tabular}}}%
\enspace
\scalebox{0.8}{
\subfloat[$\mathtt{shake}$]{%
\begin{tabular}{c|r r}
k & Support & Top 3 Words \\
\hline
1 & 0.2607 & shakespeare play time \\
\hline
2 & 0.1671 & family bad sentence \\
\hline
3 & 0.1185 & romeo juliet scene \\
\hline
4 & 0.1009 & time play text \\
\hline
5 & 0.0528 & love night dream \\
\hline
\end{tabular}}}%

\scalebox{0.8}{
\subfloat[$\mathtt{f19}$]{%
\begin{tabular}{c|r r}
k & Support & Top 3 Words \\
\hline
1 & 0.1108 & readme want fix \\
\hline
2 & 0.0822 & standard test sample \\
\hline
3 & 0.0765 & dataset issue \\
\hline
4 & 0.0746 & https pip install \\
\hline
5 & 0.0688 & file git ngrams  \\
\hline
\end{tabular}}}%
\enspace
\scalebox{0.8}{
\subfloat[$\mathtt{s20}$]{%
\begin{tabular}{c|r r}
k & Support & Top 3 Words \\
\hline
1 & 0.1369 & data correct question \\
\hline
2 & 0.0968 & true points array \\
\hline
3 & 0.0787 & test case import \\
\hline
4 & 0.0762 & error redirect prefix \\
\hline
5 & 0.0615 & point report fine \\
\hline
\end{tabular}}}%
\caption{ Summary of the top five topics extracted by LDA for each online discussion forum. For each course, the topics tend to be reasonably disjoint, with the exception of common words}
\label{top3_words}
\vspace{-0.15in}
\end{table}

\begin{figure*}[t]
    \centering
    \subfloat[\texttt{Ja}\label{subfig-1}]{%
      \includegraphics[width=0.2\textwidth]{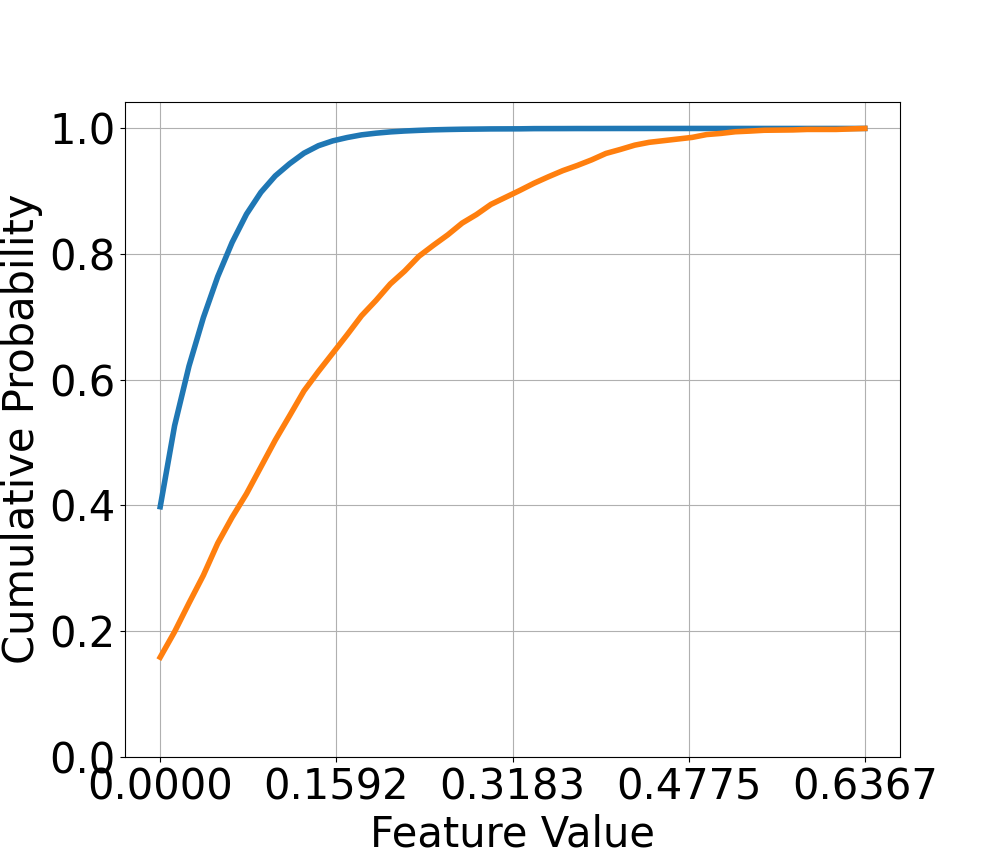}
    }
    \subfloat[\texttt{Ad}\label{subfig-2}]{%
      \includegraphics[width=0.2\textwidth]{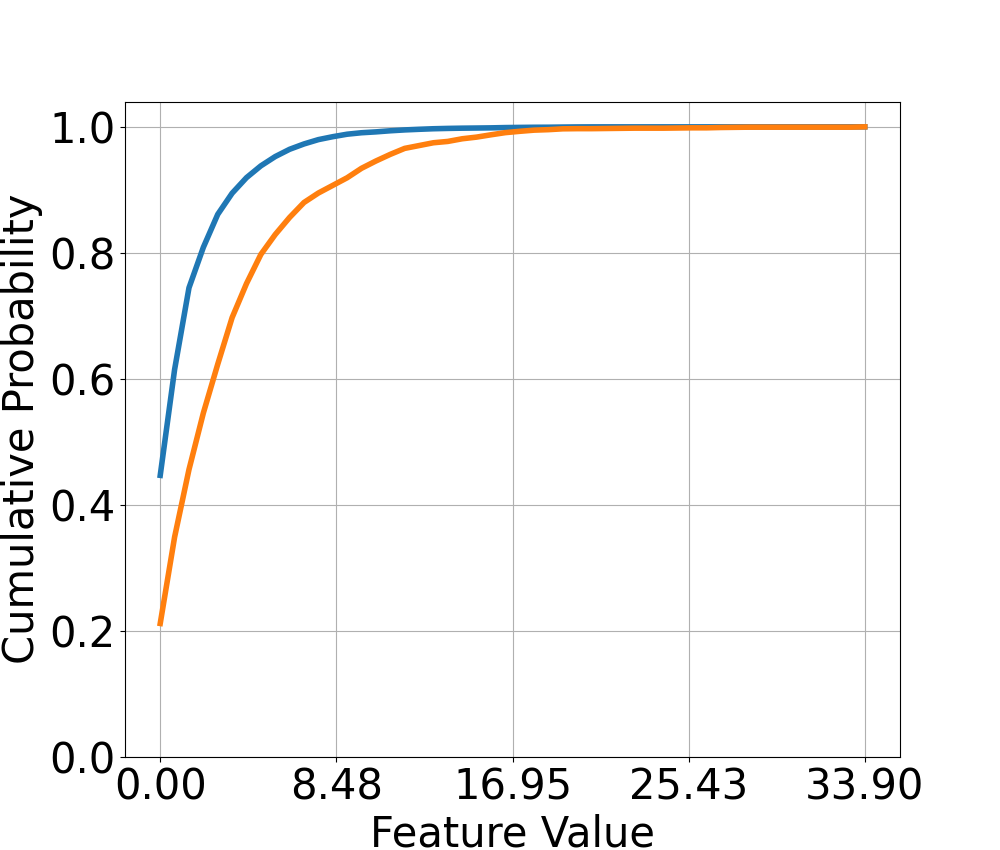}
    }
    \subfloat[\texttt{Re}\label{subfig-3}]{%
      \includegraphics[width=0.2\textwidth]{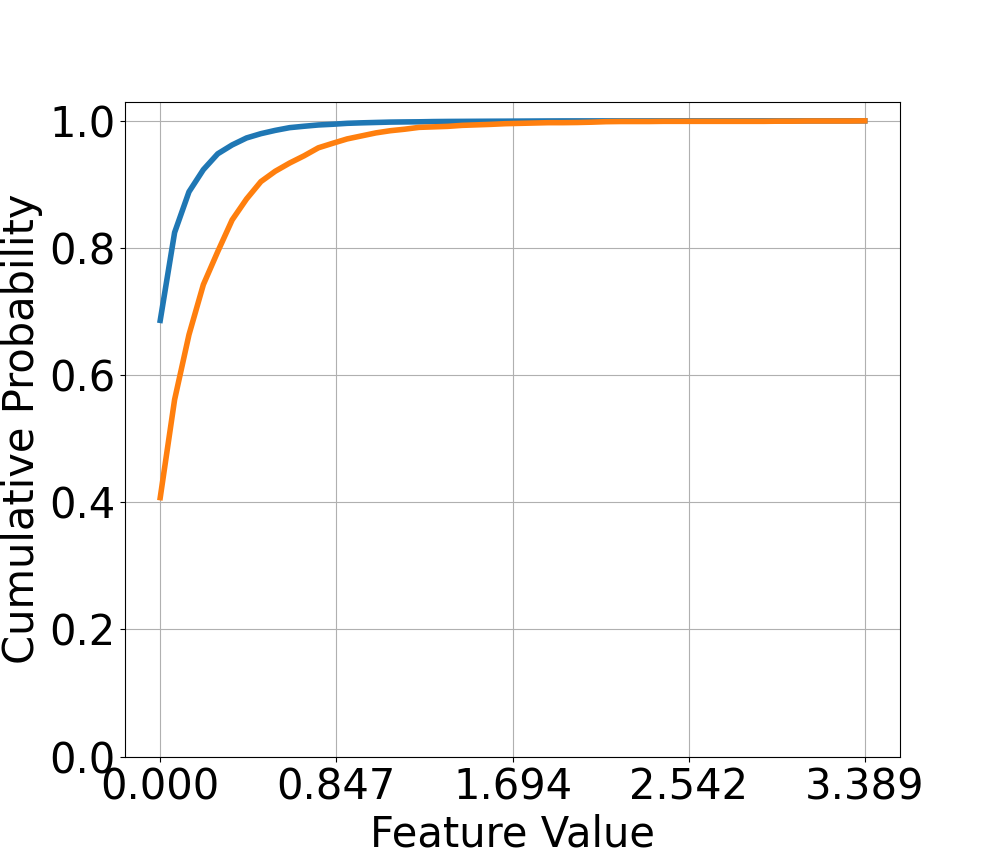}
    }
    \subfloat[\texttt{Pr}\label{subfig-4}]{%
      \includegraphics[width=0.2\textwidth]{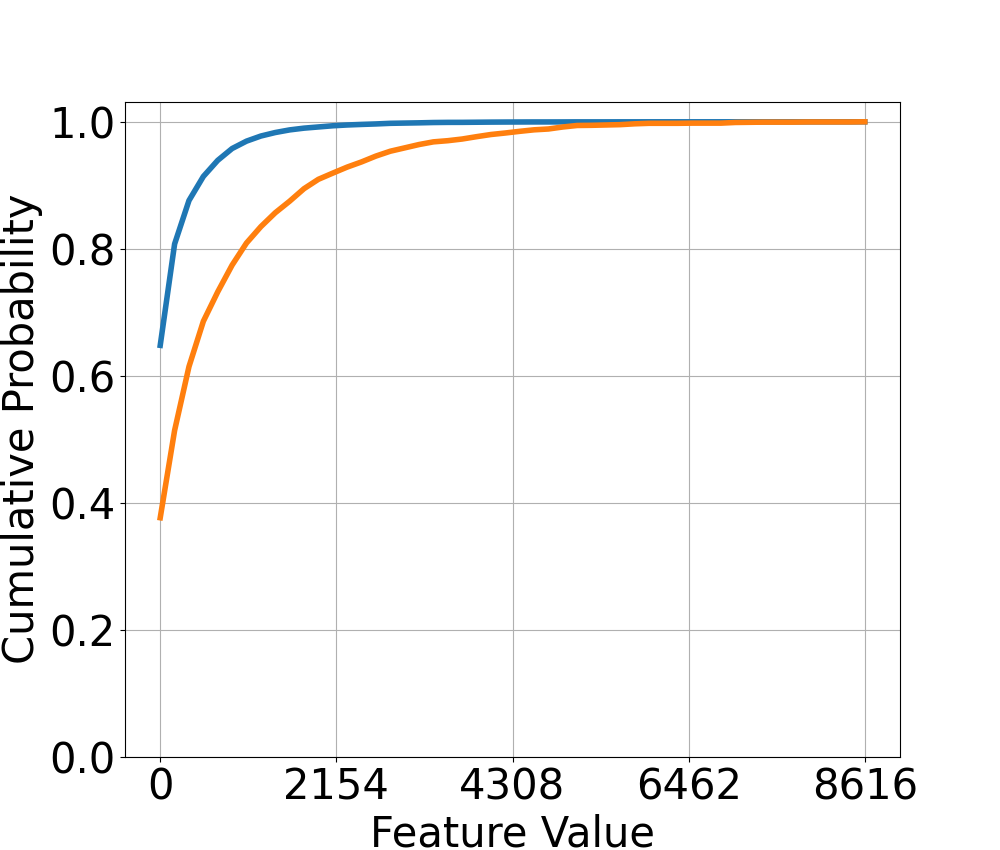}
    }
    \subfloat[\texttt{Np}\label{subfig-5}]{%
      \includegraphics[width=0.2\textwidth]{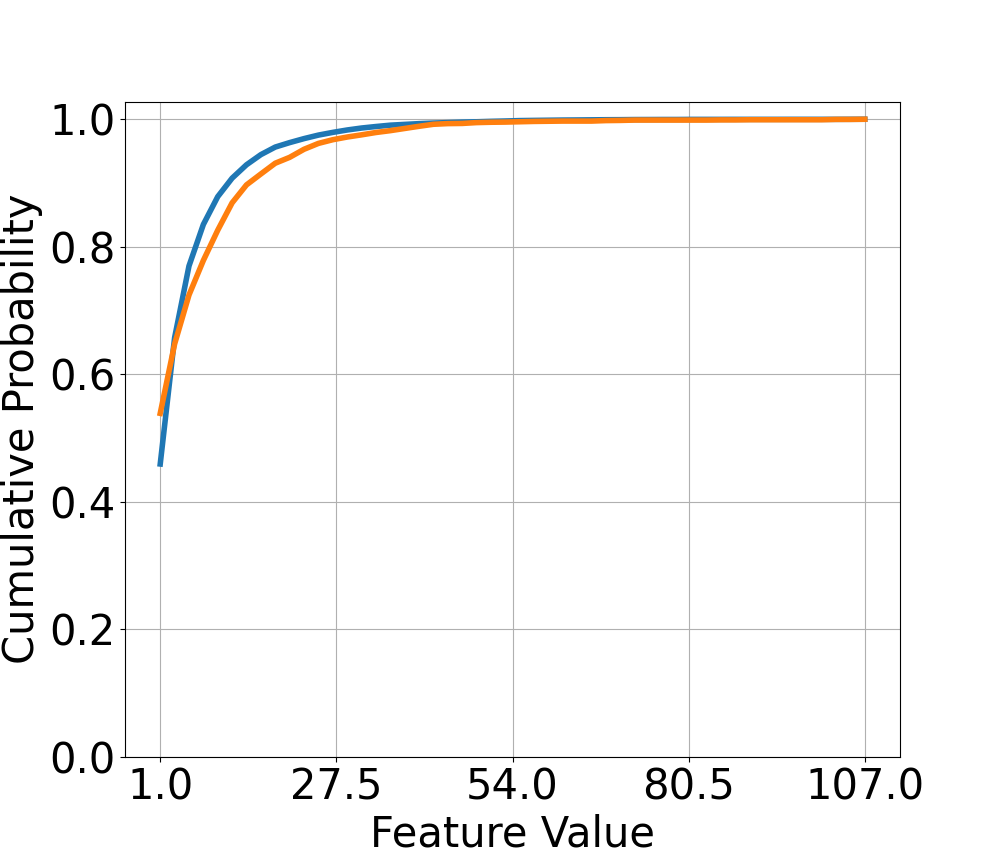}
    }
    
    \subfloat[\texttt{Lp}\label{subfig-6}]{%
      \includegraphics[width=0.2\textwidth]{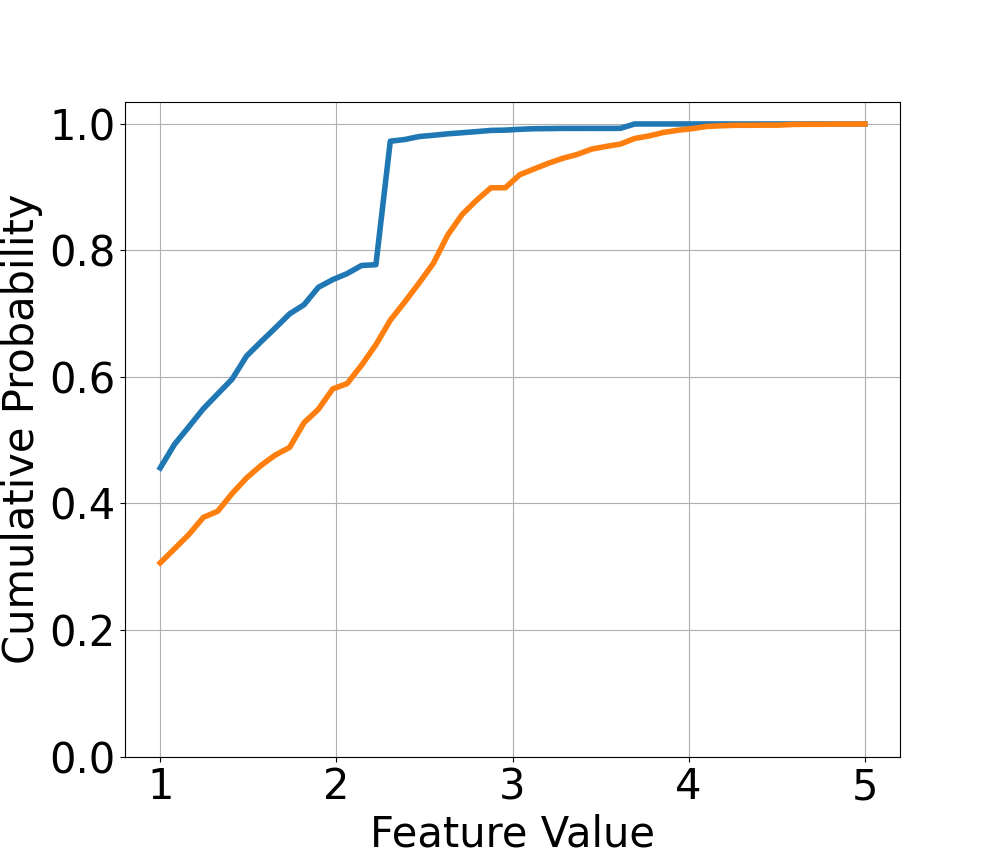}
    }
    \subfloat[\texttt{To}\label{subfig-7}]{%
      \includegraphics[width=0.2\textwidth]{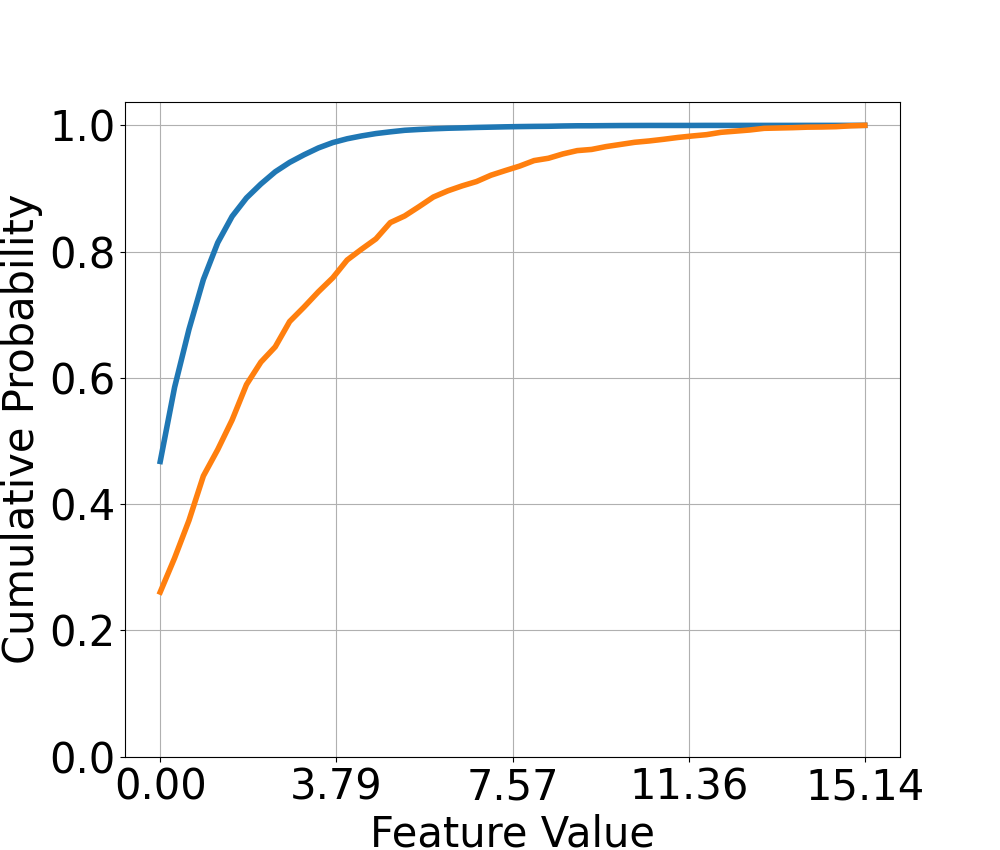}
    }
        \caption{Cumulative distribution functions (CDFs) for each of the seven feature vectors from $\mathtt{s20}$. CDFs of non-formed links are marked in blue, and CDFs of formed links are shown in orange. These demonstrate that there is (a) an observable difference in distribution between the two populations for each feature and (b) an inverse relationship between number of shortest paths and shortest path length.}
        \label{cdfs}
        \vspace{-0.15in}
\end{figure*}

\subsubsection{Deep Learning Classifiers}
\label{sec:deeplearning}

One potential limitation of linear classifiers is their small parameter space, which prevents learning intricate non-linear relationships between input features extracted from an SLN. GraphSAGE GNNs aim to address this challenge, but they lose the ability to model explicit features between node pairs. To mitigate each of these shortcomings, we propose a deep learning approach on specifically engineered features in which various characteristics of $(u, v)$ (e.g., spatial and time-varying properties) are expected to be learned for stronger prediction performance.

Specifically, we propose five deep architectures for link prediction: the Bayesian neural network (BNN), the fully connected neural network (FCNN), the convolutional neural network (CNN), the recurrent neural network (RNN), and the convolutional recurrent neural network (CRNN). Each model (excluding the Bayesian Neural Network) applies the Rectified Linear Unit (ReLU) activation function, given by $\sigma(a) = \text{max}\{0, a\}$, in its hidden layers followed by a two-unit output layer, which applies the softmax activation function, which allows for a probabilistic interpretation of link formation for a learner pair $(u, v)$. The model architecture for each of our considered models are discussed below. The hyper-parameter selection of each model was empirically determined to best fit the diverse datasets utilized in Sec. \ref{sec:modelevaluation}. 

\textbf{Bayesian Neural Network (BNN):} The Bayesian Network (BNet) model \cite{bayes} defines the probability density of latent variable $\mathbf{z}_{uv}$ as a Gaussian:
\begin{equation}
    P(\mathbf{z}_{uv}|\mathbf{e}_{uv}) = \mathcal{N}(\mathbf{w}^{T}\mathbf{e}_{uv}, \sigma^{2}),
\end{equation}
where $\mathbf{w}$ is the weight vector and $\sigma^{2}$ is the variance, both to be estimated when the model is trained. From this, $y_{uv}$ is estimated according to
\begin{equation}
    P(y_{uv} = 1|\mathbf{z}_{uv}) = \sigma(\pmb{\phi}^{T}\mathbf{z}_{uv} + b),
\end{equation}
where $\pmb{\phi}$ and $b$ are a vector and scalar, respectively, to be estimated during training, and $\sigma(\cdot)$ is the logistic sigmoid function given by $\sigma(\cdot) = 1 / (1 + e^{-(\cdot)})$. 

Our BNN architecture is composed of a hidden layer encoding the latent variable $\mathbf{z}_{uv}$. This hidden layer has 10 units, each represents a normal distribution with weight $\mathbf{w_i}$ and variance $\sigma^{2}$. Following this hidden layer is a dense output layer with softmax activation function given in \cite{bayes}.

\textbf{Fully Connected Neural Network (FCNN):} FCNNs are considered a higher dimensional non-linear extension of link classifiers. Such models can potentially represent more sophisticated non-linear relationships for better link prediction. Our fully connected multi-layer artificial neural network is composed of two hidden layers each containing 128 units. 

\textbf{Convolutional Neural Network (CNN):}
In addition to FCNN models, we also consider deep convolutional neural networks (CNNs), which in addition to providing a large parameter space for learning, capture spatial characteristics between features for each learning pair $(u, v)$. In the domain of link prediction, capturing spatial correlations between signal features is especially important since the majority of features (e.g., $\mathbf{b}_{uv}$ and $\mathbf{a}_{uv}$) are extracted from the topology of the SLN graph. 
Our proposed CNN for link prediction is composed of two convolutional layers with 64 $3 \times 1$ feature maps and 32 $2 \times 1$ feature maps, respectively, followed by a 32-unit fully connected layer.

\textbf{Recurrent Neural Network (RNN):} BNNs, FCNNs and CNNs, as well as linear classifiers, do not explicitly model the evolution of latent space variables over time based on $\mathbf{e}_{uv}$. This could potentially provide useful information for modeling an SLN, particularly so that the predictor could respond to sudden changes in the input relative to the prior state. This may occur, for example, when the topic of the course shifts, which could be reflected in a sudden change in $c_{uv}$.

To address this challenge, we consider a long-short-term memory (LSTM) based RNN with input $\mathbf{d}_{uv} = [\mathbf{e}_{uv}, \mathbf{h}_{uv}(i-1)]^{T}$, where $\mathbf{h}_{uv}(0) = 0$ and $\mathbf{h}_{uv}(i-1)$ is the output vector from the previous time. We then define the interaction gate, relationship gain gate, and relationship fading gate vectors at each time interval, $i$, as 
\begin{equation}
    \mathbf{g}_{uv}(i) = \psi(\mathbf{W}_{g}\mathbf{d}_{uv}(i) + \mathbf{b}_{g}),
\end{equation}
\begin{equation}
    \mathbf{i}_{uv}(i) = \sigma(\mathbf{W}_{i}\mathbf{d}_{uv}(i) + \mathbf{b}_{i}),
\end{equation}
\begin{equation} 
    \mathbf{f}_{uv}(i) = \sigma(\mathbf{W}_{f}\mathbf{d}_{uv}(i) + \mathbf{b}_{f}), 
\end{equation}
respectively. Here, $\psi(\cdot)$ and $\sigma(\cdot)$ are the tanh and sigmoid functions, respectively, and the matrices $\mathbf{W}_{g}$, $\mathbf{W}_{i}$, and $\mathbf{W}_{f}$ as well as the vectors $\mathbf{b}_{g}$, $\mathbf{b}_{i}$, and $\mathbf{b}_{f}$ contain parameters that are estimated during the model training procedure. 
Formally, the latent cell state, $\mathbf{z}_{uv}(i)$, is updated as
\begin{equation}
    \mathbf{z}_{uv} = \mathbf{g}_{uv}(i) \odot \mathbf{i}_{uv}(i) + \mathbf{z}_{uv}(i-1) \odot \mathbf{f}_{uv}(i),
\end{equation}
where $\odot$ denotes element-wise matrix multiplication. An output gate, $\mathbf{o}_{uv}(i)$, is then used to determine the factor to which each element of $\mathbf{z}_{uv}(i)$ should be used in the definition of $\mathbf{h}_{uv}(i)$:
\begin{equation}
    \mathbf{o}_{uv}(i) = \sigma(\mathbf{w}_{o}\mathbf{d}_{uv}(i) + \mathbf{b}_{o}), \mathbf{h}_{uv}(i) = \sigma(\mathbf{z}_{uv}(i) \odot \mathbf{o}_{uv}(i)). 
\end{equation}
With this, $y_{uv}(i)$ is estimated as 
\begin{equation}
    P(y_{uv}(i) = 1 | \mathbf{z}_{uv}(i)) = \sigma(\mathbf{h}_{1}(i)),
\end{equation}
where $\mathbf{h}_{1}(i)$ is the first element of $\mathbf{h}(i)$. Our implemented RNN is composed of 64-cell LSTM layer followed by 128-unit fully connected layer. 

\textbf{Convolutional Recurrent Neural Network (CRNN):} Convolutional recurrent neural networks contain both convolutional layers and recurrent LSTM layers. Although such models are typically computationally costly to train, they capture both spatial and time-varying correlations between learner pair feature vectors, thus providing the advantages of high parameter deep learning models with CNNs and RNNs. Our proposed CRNN architecture consists of two convolutional layers, containing $64$ $3 \times 1$ and $32$ $2 \times 1$ feature maps respectively, followed by a $32$-cell LSTM layer, and a $32$ unit fully connected layer.


\subsubsection{Deep Learning Parameter Training}
\label{sec:dlparametertraining}
We train each deep learning algorithm using the Adam optimizer as well as the categorical cross entropy loss function, which for our link prediction setup is given by 
\begin{equation}
    \mathcal{L} =  -\frac{1}{N}\sum\limits_{n=1}^{N}\sum\limits_{j=1}^{2} y_{j} \text{log}(\hat{y_{j}}),
\end{equation}
where $N$ is the total number of samples being used to calculate the loss and $\hat{y}$ is the probability of link formation. 
Each model uses a batch size of 64 as well as a learning rate of 0.001. Finally, each model is trained using 300 epochs, which is sufficient for convergence on each dataset but simultaneously allows for convergence at slightly different optima, resulting in robust and reliable evaluation when used with k-fold cross validation as further discussed in Sec. \ref{sec:modelevaluationproc}.

\section{Link Prediction Evaluation}
\label{sec:modelevaluation}

In this section, we begin by describing our considered courses along with their corresponding datasets (Sec. \ref{sec:datasets}) as well as our model evaluation procedure (Sec \ref{sec:modelevaluationproc}). We then evaluate our framework's performance for predicting link formation (Sec. \ref{sec:LPE}) and examine the time-accuracy of our prediction model (Sec. \ref{sec:TAC}).
\vspace{-0.1cm}
\subsection{Datasets}
\label{sec:datasets}

\begin{figure*}[t]
\begin{center}
\includegraphics[width=0.9\textwidth]{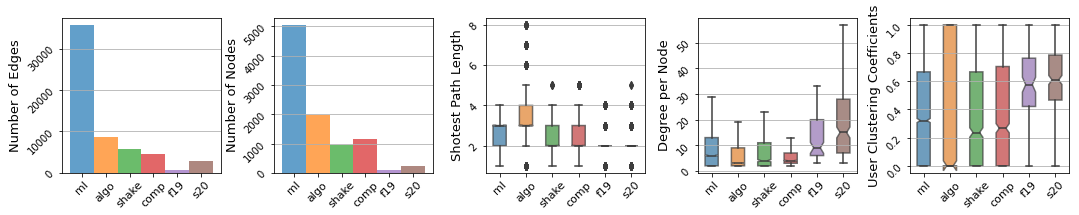}
\caption{Social network graph metrics on our datasets. We see the largest distinction in characteristics between the four MOOC courses and the two Purdue courses.}
\label{graph_metrics}
\end{center}
\end{figure*}

    
    

We consider the SLNs formed in six courses: four Coursera-based MOOC courses and two traditional courses offered at Purdue University. The four MOOC courses -- ``Machine Learning” ($\mathtt{ml}$), ``Algorithms: Design and Analysis, Part 1” ($\mathtt{algo}$), ``English Composition I” ($\mathtt{comp}$), and ``Shakespeare in Community” ($\mathtt{shake}$) -- were selected to represent a diverse set of subjects: two quantitative in nature and two in the humanities. In addition, we also consider the course ``Python for Data Science" hosted through Purdue University over two semesters: ``Fall 2019" ($\mathtt{f19}$) and ``Spring 2020" ($\mathtt{s20}$). The availability of data from two offerings of a single course provides a unique opportunity to evaluate behavior in a single course over multiple semesters. The $\mathtt{s20}$ dataset is of particular interest because of its relation with the COVID-19 pandemic. Specifically, this course was held in-person from January - March, allowing students to begin forming in-person links, which carried into their relationship in the course's SLN. However, with the pandemic forcing a transition to fully online learning, link formation between students became completely dependent on discussion forum communication. The inclusion of the $\mathtt{f19}$ and $\mathtt{s20}$ datasets, which differ both in size and in format, demonstrate our framework's broad applicability to different online course formats in dynamic environments. Table \ref{course_stats} shows detailed metrics of the six considered datasets. 

Fig. \ref{graph_metrics} summarizes the graph topology at the termination of each course under evaluation in terms of five social network metrics: number of nodes, number of edges, shortest path lengths (i.e., the $\texttt{Lp}_{uv}$ feature), degree per node, and user clustering coefficients. The diverse nature of each course is evident from each of the shown metrics and particularly from the varying number of edges and nodes. We observe the largest differences between the Purdue $\mathtt{f19}$ and $\mathtt{s20}$ courses versus the MOOC courses: the $\mathtt{f19}$ and $\mathtt{s20}$ courses are significantly smaller in nodes/edges and also have significantly larger degree per node and clustering coefficients. We also observe the difference in both the number of edges and the average degree per node between the $\mathtt{f19}$ and $\mathtt{s20}$ courses, which demonstrates the increase in student utilization of discussion forums in the absence of in-person instruction.


Next, we describe the SLNs in terms of the features in Sec. \ref{sec:featureengineering}. We make several observations on associations with link formation within and across datasets before evaluating the link-prediction portion of our proposed framework. 

\subsubsection{Data Preparation}
\label{sec:preprocessing}

To obtain a representative set of student behavior from a course, and to ensure that data gathered from each source is uniformly formatted, we filter each considered dataset. Specifically, we remove the instructors from the list of learners and remove all links formed between learners and instructors, since we are interested in developing models targeted towards peer-to-peer interaction, with the goal of requiring less direct instructor intervention. Furthermore, interactions before the beginning of a course are removed; only links formed during a course are considered. Both course-hosting sites offer an option for full anonymity to learners -- posts made with anonymity are ignored, as we cannot make meaningful connections with unknown users. Enrolled learners who did not access the forum (i.e., an empty adjacency matrix), are not considered to remove confusion -- a lack of behavior excludes a helpful metric for predicting future behavior. Such students would likely benefit from more traditional intervention. After filtering, less than 2\% of the learner pairs in each dataset demonstrated a formed link. This underscores an extreme sparsity of learner pairs for link prediction; the methodology applied to avoid overfitting will be discussed further in Section \ref{sec:modelevaluationproc}.

\subsubsection{Topic extraction}\label{sec:topicextraction} To obtain the post similarities $c_{uv}(i)$, we must first extract the topics, $\mathcal{K}$, and distributions for each post according to the LDA algorithm discussed in Sec. \ref{sec:featureengineering}. Prior to building the dictionary of topics, all URLs, punctuations, and stopwords are removed from each post’s text and all words are stemmed. Table \ref{top3_words} summarizes the topic extraction results for each dataset using $|\mathcal{K}|$ = 20 topics; the top three words shown are from the five topics that have the highest supports across posts. We find that $|\mathcal{K}| = 20$ produces a set of topics that have reasonably large supports across posts while retaining granular information, i.e., able to convey differences between student posts. In our manual inspection, larger values of $|\mathcal{K}|$ lacked the support to generate informative features, while smaller values of $|\mathcal{K}|$ resulted in too much intersection between topics for a good understanding of content.

\begin{table*}[!ht]
\centering
\begin{tabular}{c c c c c c c c} 
 \hline
 \multicolumn{2}{c}{Model} & $\mathtt{ml}$ & $\mathtt{algo}$ & $\mathtt{shake}$ & $\mathtt{comp}$ & $\mathtt{f19}$ & $\mathtt{s20}$\\
  \hline
 \hline
 \multirow{2}{1.5cm}{$\mathtt{Re}$} & AUC & 0.5005 $\pm$ 0.0004 & 0.5188 $\pm$ 0.0322 & 0.5061 $\pm$ 0.0034 & 0.5167 $\pm$ 0.0266 & 0.5689 $\pm$ 0.0401 & 0.5238 $\pm$ 0.0121\\ 
  & ACC & 0.5995 $\pm$ 0.0054 & 0.8338 $\pm$ 0.0104 & 0.8296 $\pm$ 0.0073 &  0.8349 $\pm$ 0.0082 & 0.9524 $\pm$ 0.0057 & 0.9599 $\pm$ 0.0020\\
  \hline
 \multirow{2}{1.5cm}{$\mathtt{BNet}$} & AUC & 0.9053 $\pm$ 0.0106 & 0.9488 $\pm$ 0.0058 & 0.8603 $\pm$ 0.0095 & 0.8684 $\pm$ 0.0116 & 0.7413 $\pm$ 0.0546 & 0.7495 $\pm$ 0.0269\\ 
  & ACC & 0.9175 $\pm$ 0.0066 & 0.9805 $\pm$ 0.0019 & 0.9472 $\pm$ 0.0035 & 0.9492 $\pm$ 0.0026 & 0.9600 $\pm$ 0.0053 & 0.9672 $\pm$ 0.0013\\
    \hline
 \multirow{2}{1.5cm}{$\mathtt{FCNN}$} & AUC & 0.9766 $\pm$ 0.0033 & 0.9706 $\pm$ 0.0039 & 0.9670 $\pm$ 0.0059 & 0.9714 $\pm$ 0.0084 & \textbf{0.8991 $\pm$ 0.0367} & 0.8844 $\pm$ 0.0330 \\ 
  & ACC & 0.9782 $\pm$ 0.0027 & 0.9871 $\pm$ 0.0029 & 0.9853 $\pm$ 0.0019 & 0.9850 $\pm$ 0.0022 & \textbf{0.9688 $\pm$ 0.0037} & {0.9729 $\pm$ 0.0022}\\
    \hline
 \multirow{2}{1.5cm}{$\mathtt{SVM}$} & AUC & 0.9122 $\pm$ 0.0027 & 0.9523 $\pm$ 0.0050 & 0.8982 $\pm$ 0.0071 & 0.8618 $\pm$ 0.0071 & 0.8437 $\pm$ 0.0343 & 0.8203 $\pm$ 0.0113\\ 
  & ACC & 0.9137 $\pm$ 0.0026 & 0.9755 $\pm$ 0.0035 & 0.9608 $\pm$ 0.0031 & 0.9462 $\pm$ 0.0022 & 0.9670 $\pm$ 0.0040 & 0.9700 $\pm$ 0.0015\\
    \hline
 \multirow{2}{1.5cm}{$\mathtt{LinDA}$} & AUC & 0.8486 $\pm$ 0.0056 & 0.8361 $\pm$ 0.0064 & 0.7521 $\pm$ 0.0116 & 0.7331 $\pm$ 0.0123 & 0.6940 $\pm$ 0.0146 & 0.6692 $\pm$ 0.0205\\ 
  & ACC & 0.8674 $\pm$ 0.0051 & 0.9425 $\pm$ 0.0018 & 0.9117 $\pm$ 0.0050 & 0.9084 $\pm$ 0.0056 & 0.9582 $\pm$ 0.0046 & 0.9620 $\pm$ 0.0026\\
 \hline
 \multirow{2}{1.5cm}{$\mathtt{RNN}$} & AUC & \textbf{0.9880 $\pm$ 0.0011} & \textbf{0.9808 $\pm$ 0.0026} & \textbf{0.9807 $\pm$ 0.0054} & \textbf{0.9770 $\pm$ 0.0071} & {0.8304 $\pm$ 0.0373} & {0.8329 $\pm$ 0.0349}\\ 
  & ACC & \textbf{0.9890 $\pm$ 0.0010} & \textbf{0.9902 $\pm$ 0.0013} & \textbf{0.9906 $\pm$ 0.0019} & \textbf{0.9877 $\pm$ 0.0030} & {0.9653 $\pm$ 0.0040} & {0.9710 $\pm$ 0.0024}\\
  \hline
   \multirow{2}{1.5cm}{$\mathtt{CNN}$} & AUC & \textbf{0.9881 $\pm$ 0.0019} & \textbf{0.9817 $\pm$ 0.0029} & \textbf{0.9754 $\pm$ 0.0057} & \textbf{0.9763 $\pm$ 0.0055} & \textbf{0.9187 $\pm$ 0.0318} & \textbf{0.9221 $\pm$ 0.0169}\\
  & ACC & \textbf{0.9894 $\pm$ 0.0015} & \textbf{0.9916 $\pm$ 0.0009} & \textbf{0.9888 $\pm$ 0.0025} & \textbf{0.9882 $\pm$ 0.0022} & \textbf{0.9711 $\pm$ 0.0033} & \textbf{0.9740 $\pm$ 0.0015}\\
   \hline
   \multirow{2}{1.5cm}{$\mathtt{CRNN}$} & AUC & 0.9680 $\pm$ 0.0094 & 0.9704 $\pm$ 0.0087 & {0.9608 $\pm$ 0.0066} & \textbf{0.9725 $\pm$ 0.0070} & \textbf{0.8903 $\pm$ 0.0468} & 0.8845 $\pm$ 0.0347\\
  & ACC & 0.9713 $\pm$ 0.0090 & 0.9846 $\pm$ 0.0036 & 0.9803 $\pm$ 0.0028 & \textbf{0.9859 $\pm$ 0.0020} & \textbf{0.9705 $\pm$ 0.0016} & {0.9724 $\pm$ 0.0020}\\
 \hline
 \multirow{2}{1.5cm}{$\mathtt{GNN}$} & AUC & \textbf{0.9969 $\pm$ 0.0014} & \textbf{0.9989 $\pm$ 0.0007} & \textbf{0.9988 $\pm$ 0.0011} & \textbf{0.9955 $\pm$ 0.0029} & 0.7395 $\pm$ 0.0508 & 0.5628 $\pm$ 0.1157 \\
  & ACC & \textbf{0.9967 $\pm$ 0.0008} & \textbf{0.9988 $\pm$ 0.0007} & \textbf{0.9965 $\pm$ 0.0068} & \textbf{0.9958 $\pm$ 0.0019} & 0.6557 $\pm$ 0.0751 & 0.5500 $\pm$ 0.0997\\
 \hline
\end{tabular}
\caption{Performance of each considered link prediction model. The $\mathtt{CNN}$ model is among the best performing model across all six datasets with respect to the AUC and ACC metrics. All results in \textbf{bold} highlight the best performing results. We see that the $\mathtt{GNN}$ results in strong performance on the four MOOC courses while performing poorly on the two Purdue courses, indicating that GNNs are effective for link prediction in large courses whereas our method delivers strong performance in small courses as well as large.}
\label{model_perf}
\end{table*}
\vspace{-0.1cm}
\subsection{Model Evaluation Procedure}
\label{sec:modelevaluationproc}

To evaluate the models proposed in Sec. \ref{sec:SLNmodel}, we use the following metrics, training procedures, and evaluation criteria. 

\subsubsection{Metrics} We use three metrics to evaluate prediction performance. First, we compute the overall Accuracy (ACC), or the fraction of predictions over all time that are correct. For iteration $k$, it is obtained as:
\begin{equation}
    \frac{1}{|\Omega_e^k| \cdot L} \sum\limits_{(u,v)\in\Omega_e^k}\sum_{i = 1}^{L} \mathbbm{1}\{y_{uv}(i) =  \bar{y}_{uv}(i)\},
\end{equation}
where $y_{uv}(i) \in \{0, 1\}$ is the binary prediction made based on
$\tilde{y}_{uv}(i)$ and $\mathbbm{1}$ is the indicator function. Second, we compute the Area Under the ROC Curve (AUC), which assesses the tradeoff between true and false positive rates for a classifier \cite{moocperformance}. Third, we define a metric called Time Accuracy (TAC) to be the fraction of links that are predicted to form within a fixed window $w$ of when they actually form (among those that eventually form). Letting $n_{uv} = \text{min}_{i}\{y_{uv}(i) = 1\}$ be the actual time at which link $(u, v) \in \Omega_k^f$ forms and $\tilde{n}_{uv} = \text{min}_{i}\{\tilde{y}_{uv}(i) = 1\}$ the predicted time, the TAC is defined as
\begin{equation}
    \frac{1}{|\Omega_k^f|} \sum\limits_{(u,v)\in\Omega_k^f}\mathbbm{1}\{|\tilde{n}_{uv} - n_{uv}| \leq w\}
\end{equation}
for iteration $k$, where $\Omega_f^k \subset \Omega_e^k$ is the set of correctly predicted links in the test set that will eventually form. We compute the mean and standard deviation of each metric across three evaluation iterations.


\subsubsection{Training and Testing}
\label{sec:traintest}
$k$-fold cross validation is used to evaluate each predictor with $k = 10$. Following Sec. \ref{sec:datasets}, we again consider the link sets $\mathcal{G}(L)$ and $\mathcal{G}^{c}(L)$. Our objective is to train models capable of accurate link prediction despite the large class imbalance between $\mathcal{G}(L)$ and $\mathcal{G}^{c}(L)$ that will be observed during training and inference. To achieve this, we take an equal proportion of samples from both $\mathcal{G}(L)$ and $\mathcal{G}^{c}(L)$ to form each training fold, which, in turn, retains the overall class imbalance in the training set during each training iteration. The corresponding testing set of each training fold contains the same class imbalance. After each training fold, we calculate the metrics of interest on the respective testing set of the validation run. This sampling, along with the utilization of the AUC measurement, allows us to quantify the false alarm versus true positive rate, since the prediction accuracies on a poorly trained model could be very high due to the large class imbalance.

In each of the $k$ iterations, we consider a set of time intervals from which the model parameters are estimated considering each pair $(u, v) \in \Omega^{r}_{k}$, using the procedures in Sec. \ref{sec:traintest}. Then, for each $(u, v) \in \Omega^{e}_{k}$, the inputs are used to make a prediction $\tilde{y}_{uv}(i) \in [0, 1]$ of the link state $y_{uv}(i)$.


\subsection{Link Prediction Evaluation}
\label{sec:LPE}


Table \ref{model_perf} gives the overall performance of the baseline, linear, GNN, and deep learning models in terms of the AUC and ACC metrics. Overall, we see that the $\mathtt{CNN}$ consistently outperforms the other predictors for each considered dataset. In addition, the $\mathtt{GNN}$ achieves strong (comparable to the $\mathtt{CNN}$) prediction performance on the four MOOC datasets, but it performs poorly on both $\mathtt{f19}$ and $\mathtt{s20}$, achieving AUCs of 0.74 and 0.56 in $\mathtt{f19}$ and $\mathtt{s20}$, respectively. This behavior is consistent with observations in prior work \cite{gnns} that GNNs require large datasets for effective generalization -- a characteristic that MOOCs are able to provide (with at least 1,000 users in each case, see Table \ref{course_stats}) whereas the Purdue courses, $\mathtt{f19}$ and $\mathtt{s20}$, are not. Our explicit feature engineered methodology paired with a CNN classifier, on the other hand, is more robust against variations in SLN course size and type in comparison to the GNN.

Of particular interest is the $\mathtt{s20}$ dataset and its performance relative to the other five datasets. Because $\mathtt{s20}$ was held partially in-person prior to the COVID-19 outbreak in March 2020, the behavior represented includes both in-person and online interactions. Furthermore, it contains a rapid change in behavior midway through the semester that models must account for. It follows from the high accuracies and AUCs demonstrated by each deep-learning model on this dataset that our prediction model can be applied to hybrid-online courses with a similar level of accuracy to fully online courses. It also suggests that our proposed model is responsive to large-scale shifts in student behavior. From Table \ref{model_perf}, we see neither the $\mathtt{GNN}$ nor the other baseline models are capable of capturing either of these desirable characteristics. As a result, we find that our proposed framework is capable of increasing both course quality and learner interactions during the pandemic; an attribute that can be leveraged to improve instruction in a post-pandemic course offering. 

Considering all courses, the $\mathtt{CNN}$ model has slightly higher performance across the metrics and datasets, reaching average AUCs between 0.92 and 0.99 and average ACCs between 0.97 and 0.99. The AUC of $\mathtt{Re}$ is nearly random, but demonstrates a high accuracy in all cases because of the large class imbalance present. Similarly, the linear classifiers demonstrate high ACC values because of the large class imbalance as well. Although the Bayesian model consistently outperforms the baseline models, the lower accuracy and AUC relative to the $\mathtt{CNN}$ and $\mathtt{CRNN}$ models confirms our hypothesis from Sec. \ref{sec:SLNmodel} that capturing spatial and temporal variance leads to improvement in the model. More specifically, the evolution of the state of an SLN between different time periods, both temporally and spatially, is important to predicting learner interactions; this aspect is effectively included in the LSTM-based $\mathtt{CRNN}$. We further observe that the $\mathtt{CNN}$ model, capturing spatial variance, and the $\mathtt{RNN}$ capturing temporal variance, each perform similarly to the $\mathtt{CRNN}$ model for several datasets. This suggests that while spatial and temporal variance both individually assist in prediction, their combined usage may not result in significant performance improvements. 

Although an accurate prediction is most informative on the efficacy of a connection between learners, recommendations may also be supported by false predictions. If a high-accuracy model falsely predicts that two users will connect, we may infer that the formation of a link between these two users would be beneficial based on model parameters. Conversely, there is a strong correlation between false negative predictions and weak links between learners, implying that the benefits of forming a connection between two such users would be trivial compared to other, more highly-weighted connections.


\begin{figure*}[t]
    \centering
    \subfloat[$\mathtt{ml}$\label{subfig-1}]{%
      \includegraphics[width=0.15\textwidth]{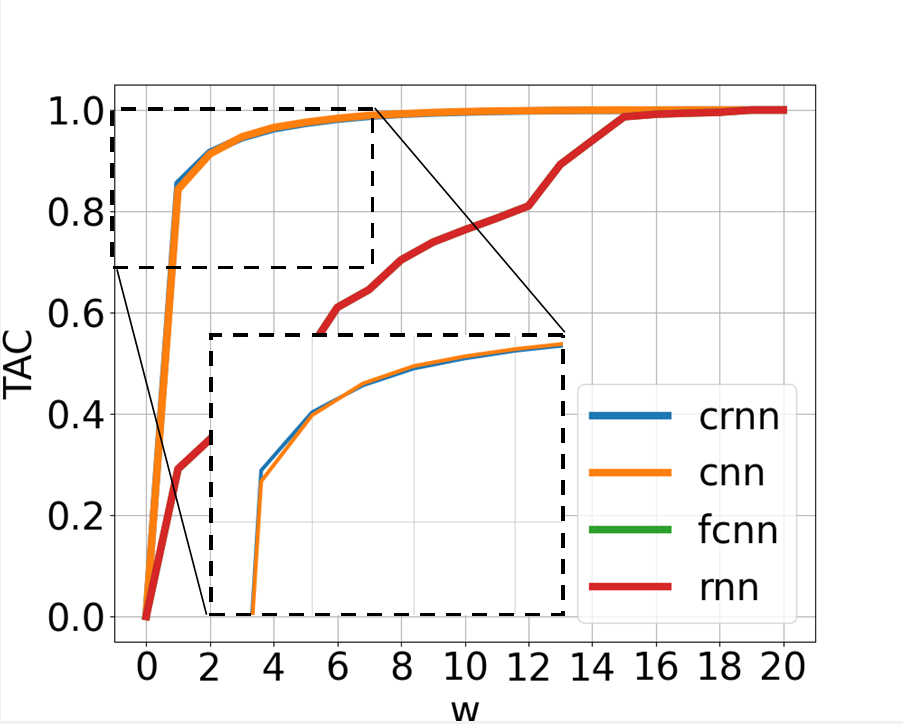}
    }
    \subfloat[$\mathtt{algo}$\label{subfig-2}]{%
      \includegraphics[width=0.15\textwidth]{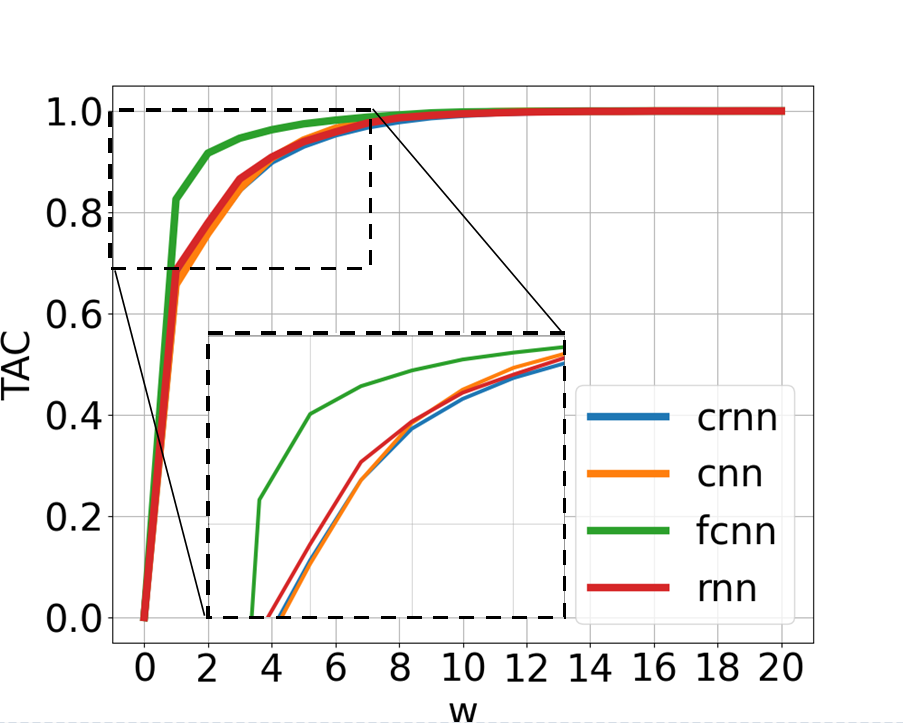}
    }
    \subfloat[$\mathtt{shake}$\label{subfig-3}]{%
      \includegraphics[width=0.15\textwidth]{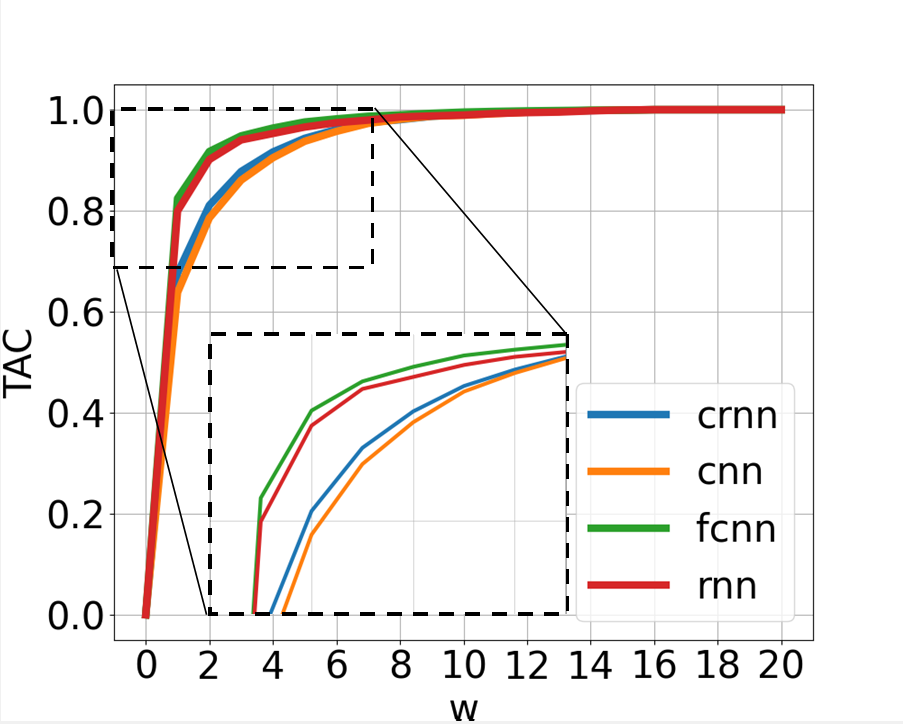}
    }
    \subfloat[$\mathtt{comp}$\label{subfig-4}]{%
      \includegraphics[width=0.15\textwidth]{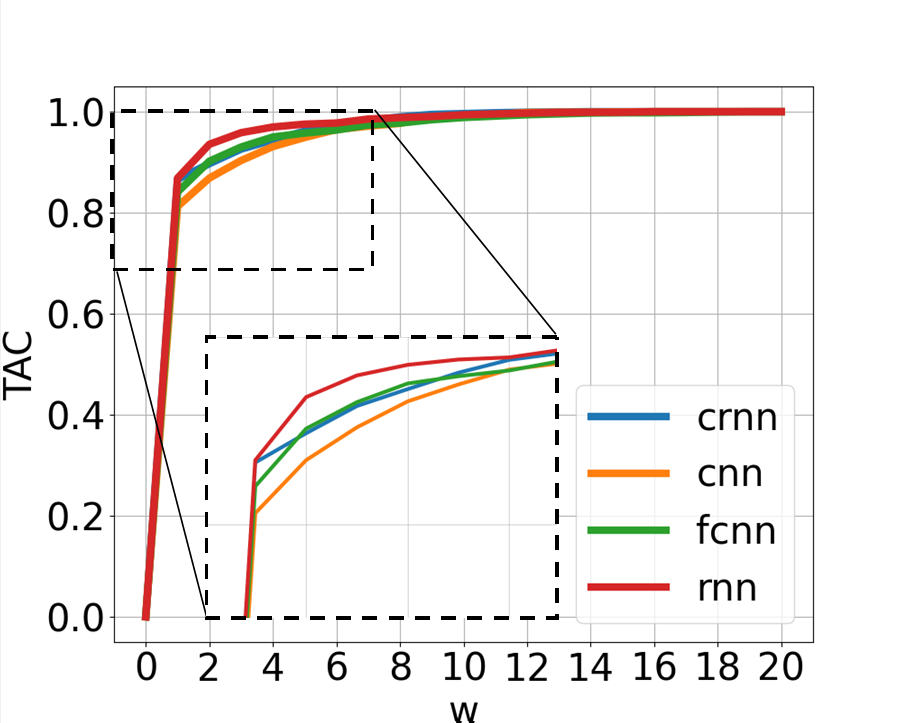}
    }
    \subfloat[$\mathtt{f19}$\label{subfig-5}]{%
      \includegraphics[width=0.15\textwidth]{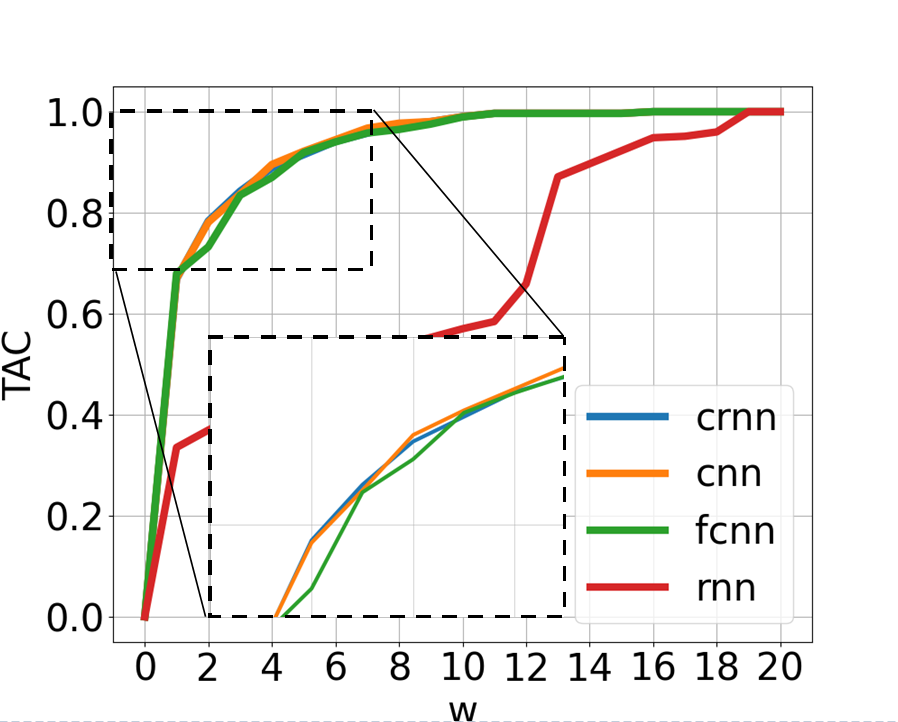}
    }
    \subfloat[$\mathtt{s20}$\label{subfig-6}]{%
      \includegraphics[width=0.15\textwidth]{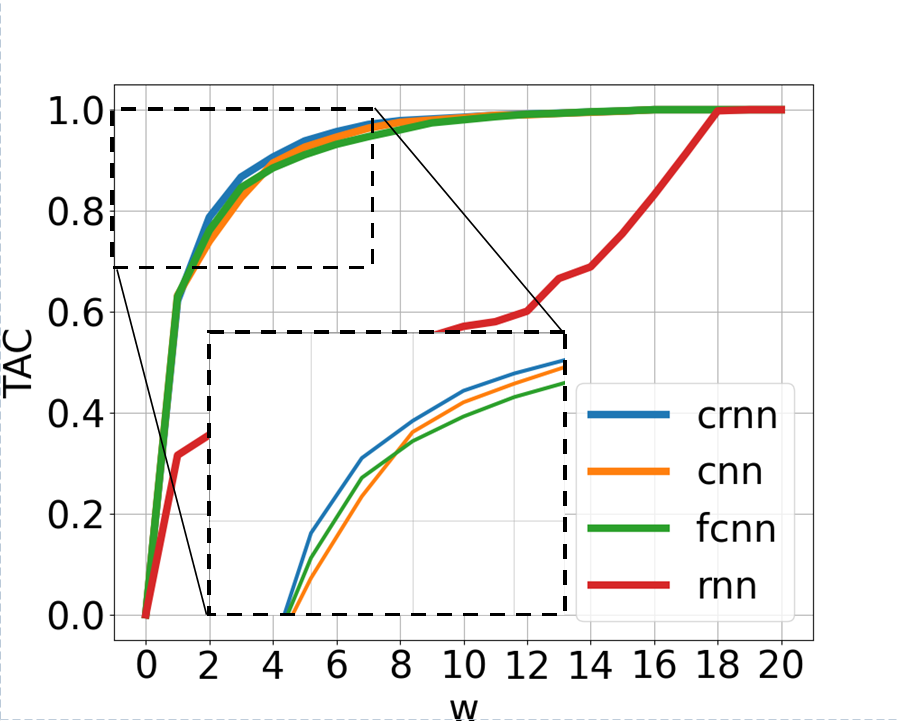}
    }
        \caption{TAC with different windows $w$. The TAC curves all exhibit sharp increases initially, indicating many links form around the time they are predicted to. The links at higher $w$, on the other hand, indicate potential for recommending early link formation and future reconnection.}
        \label{tac}
\end{figure*}

\subsection{Early Detection of Link Formation}
\label{sec:TAC}
The models proposed in Sec. \ref{sec:LPE} consider the ability to predict link formation in subsequent time intervals up until the end of the course. However, it does not consider links that will form at an earlier or later interval. These occurrences of a delay between link formation and prediction can lend additional information of importance to learners: if we can predict in advance which learners may form connections, we may encourage them to connect sooner, potentially resulting in a stronger connection or faster replies from learners expected to have delayed responses. On the other hand, if we find that a link forms much sooner than predicted by our model, this may indicate that learners would benefit from re-connecting on the current topic later in the course. 


To study these cases, we evaluate the TAC metric from Sec. \ref{sec:modelevaluationproc} for our $\mathtt{RNN}$, $\mathtt{CNN}$, $\mathtt{FCNN}$, and $\mathtt{CRNN}$ models; i.e., we measure whether links form within a given window $w$ of when they are predicted to. Note that the TAC metric was only calculated for the deep learning models, since they were consistently the best performing link formation predictors. The granular value of 20 time intervals used to generate the SLN graph model gives the predictive model access to more frequently updated features, and allows the model to respond quickly to changes in SLN behavior. Fig. \ref{tac} shows the TAC values as $w$ is increased from 0 to 20 for several of our proposed deep learning prediction models. The sharp increase of each TAC curve for small $w$ of each model -- with the exception of the RNN -- indicates that many links form close to when they are predicted to form, reinforcing our observations of model quality from other performance metrics in Sec. \ref{sec:LPE}. A window of $w = 2$, for example, is already sufficient for all six forums to reach a TAC of 0.5 or above. 

Observing Fig. \ref{subfig-5}, which represents the TAC curve of the $\mathtt{f19}$ dataset, it is clear that our TAC metric demonstrates a lower accuracy for small datasets but the performance of individual models has more variation. This is largely attributed to the smaller number of learner pairs contained in the $\mathtt{f19}$ dataset with which to train the model compared to a MOOC forum. However, with the exception of the $\mathtt{RNN}$, we can observe the same curve shape and sharp initial increase present for larger datasets, indicating that TAC is both a consistent and useful evaluation metric of model performance. We can further observe in the \texttt{ml}, \texttt{f19}, and \texttt{s20} datasets that the $\mathtt{RNN}$ model fails to correctly predict links consistently across datasets within a small interval of when they actually occur, further suggesting that spatial features play a more important role in the problem of link prediction, which is further discussed in Sec. \ref{sec:moel_arch}.

Furthermore, there are very few links with large $w$, once again reinforcing the results of other performance metrics. The small quantity of links with large $w$ in each forum present a significant opportunity to recommend early formation of links (when predictions are early) and potential times for learners to reconnect (when predictions are late). Though there is less room for change on links with smaller $w$, learners may be more willing to act on recommendations in these cases since they induce less modification to actual behavior \cite{moocopt}; after all, a learner may be reluctant to reach out to others on the basis of outdated threads or on the assumption that they will eventually collaborate.

\begin{figure*}[t]
    \centering
    \includegraphics[width=0.165\textwidth]{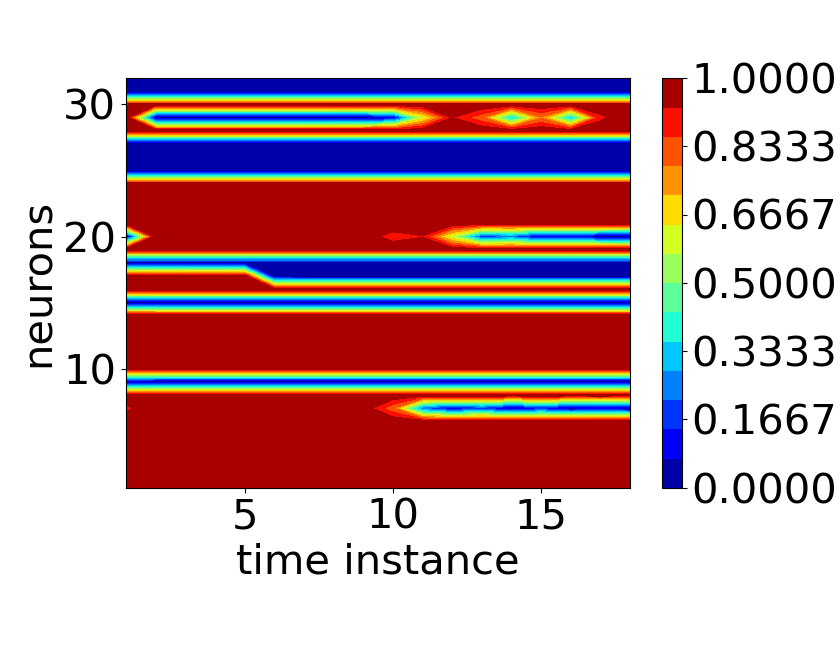}
    \includegraphics[width=0.165\textwidth]{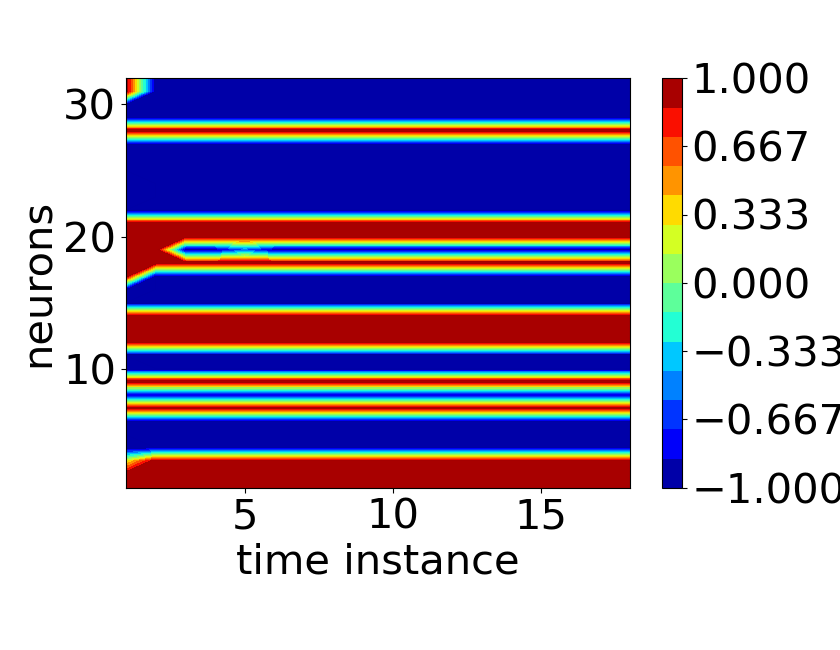}
    \includegraphics[width=0.16\textwidth]{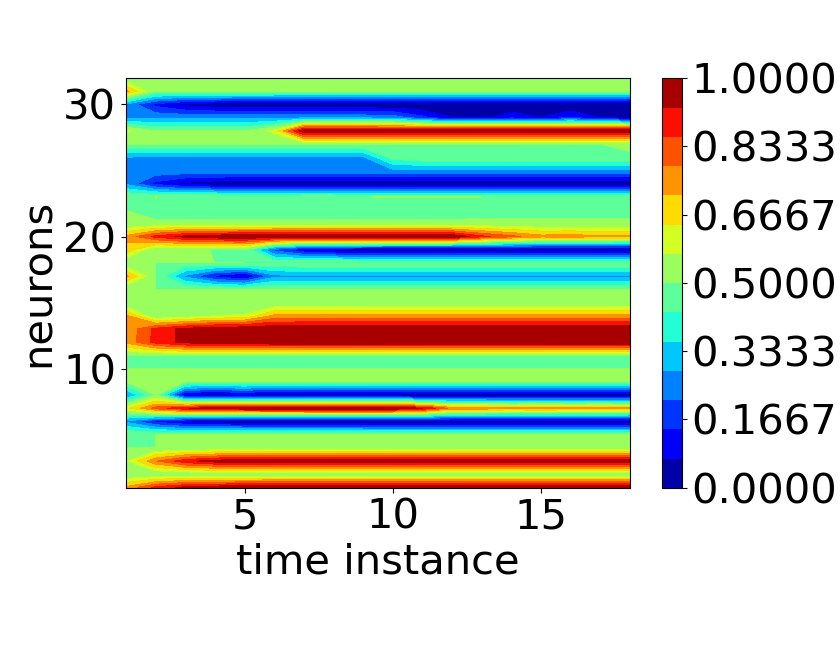}
    \includegraphics[width=0.16\textwidth]{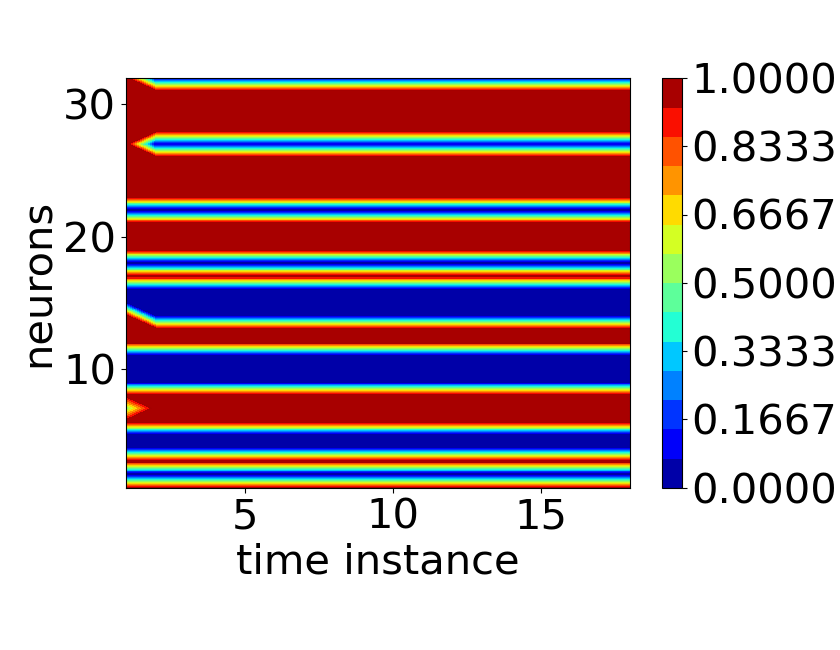}
    \includegraphics[width=0.16\textwidth]{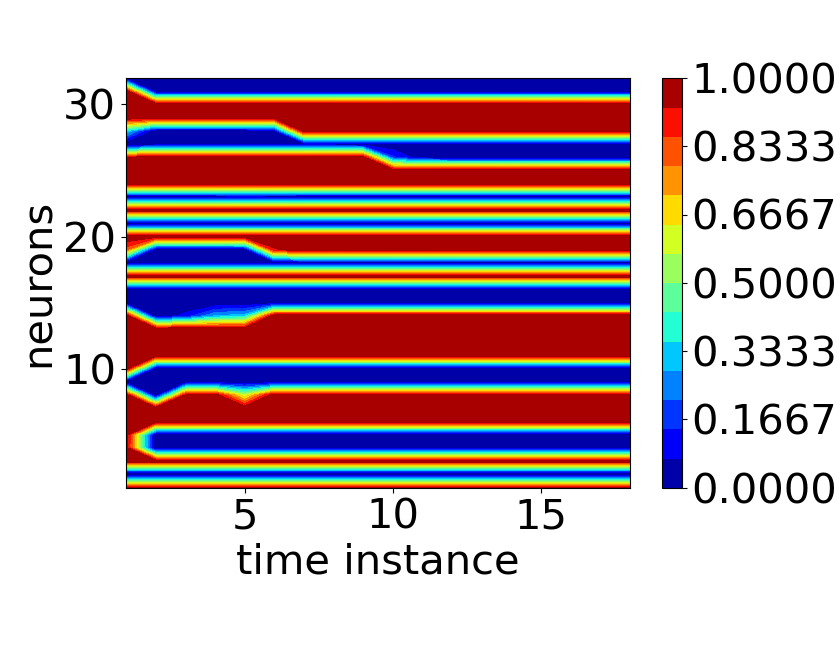}
    \includegraphics[width=0.16\textwidth]{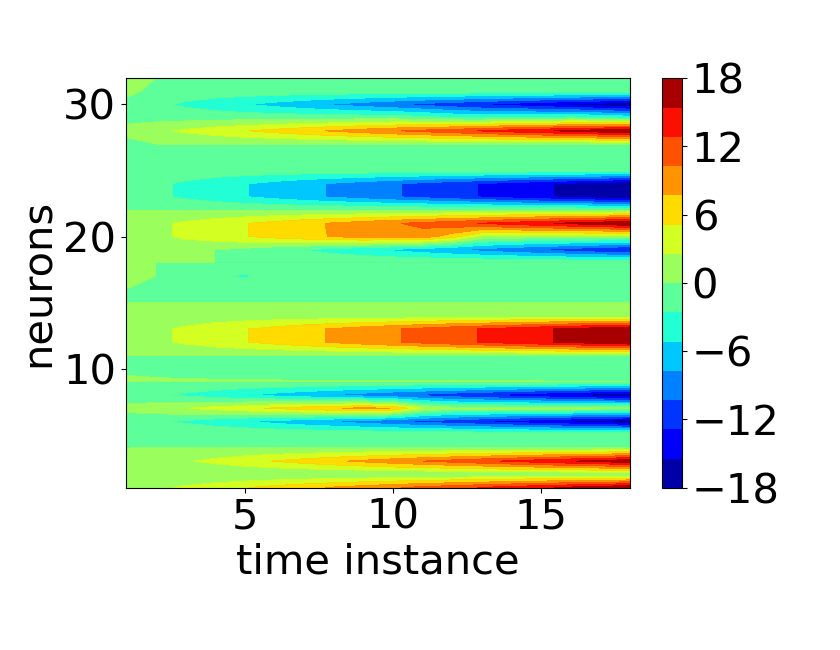}
    
    \subfloat[$\mathbf{f}_{uv}(i)$\label{subfig-1}]{%
      \includegraphics[width=0.165\textwidth]{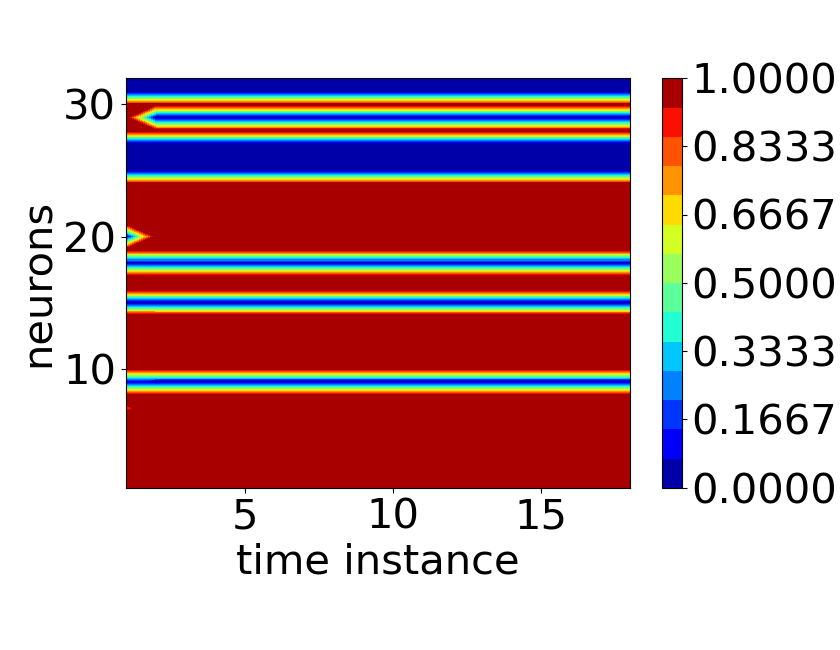}
    }
    \subfloat[$\mathbf{g}_{uv}(i)$\label{subfig-2}]{%
      \includegraphics[width=0.165\textwidth]{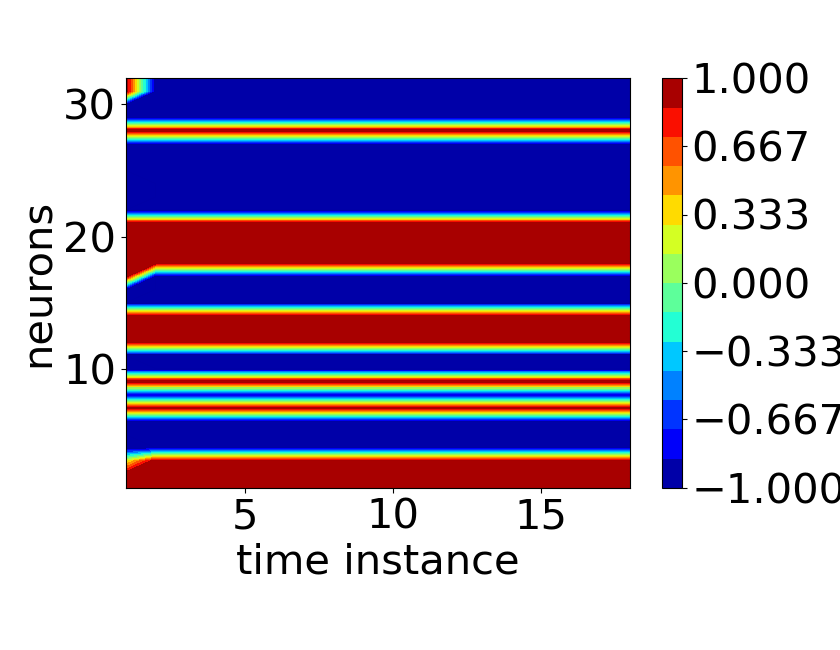}
    }
    \subfloat[$\mathbf{h}_{uv}(i)$\label{subfig-3}]{%
      \includegraphics[width=0.165\textwidth]{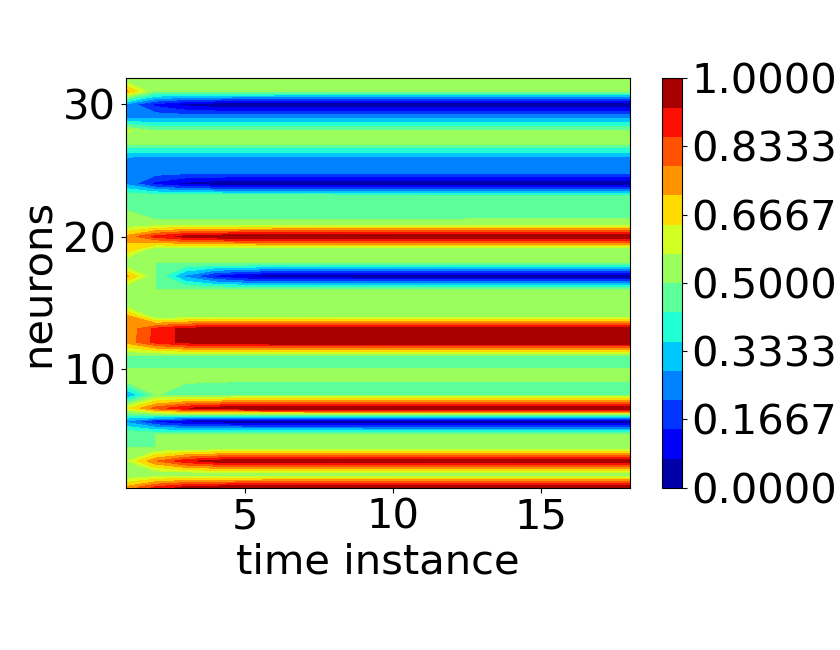}
    }
    \subfloat[$\mathbf{i}_{uv}(i)$\label{subfig-3}]{%
      \includegraphics[width=0.16\textwidth]{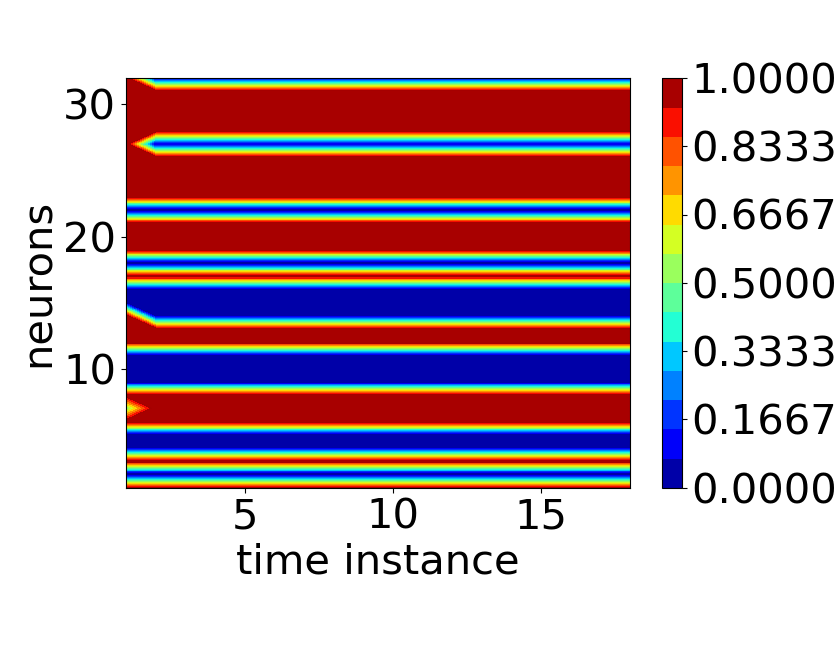}
    }
    \subfloat[$\mathbf{o}_{uv}(i)$\label{subfig-3}]{%
      \includegraphics[width=0.16\textwidth]{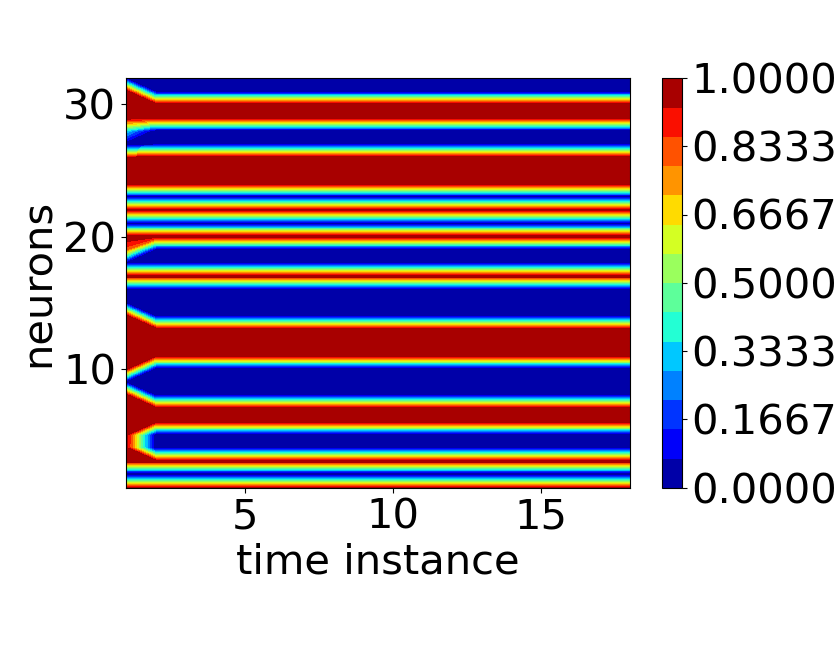}
    }
    \subfloat[$\mathbf{z}_{uv}(i)$\label{subfig-3}]{%
      \includegraphics[width=0.16\textwidth]{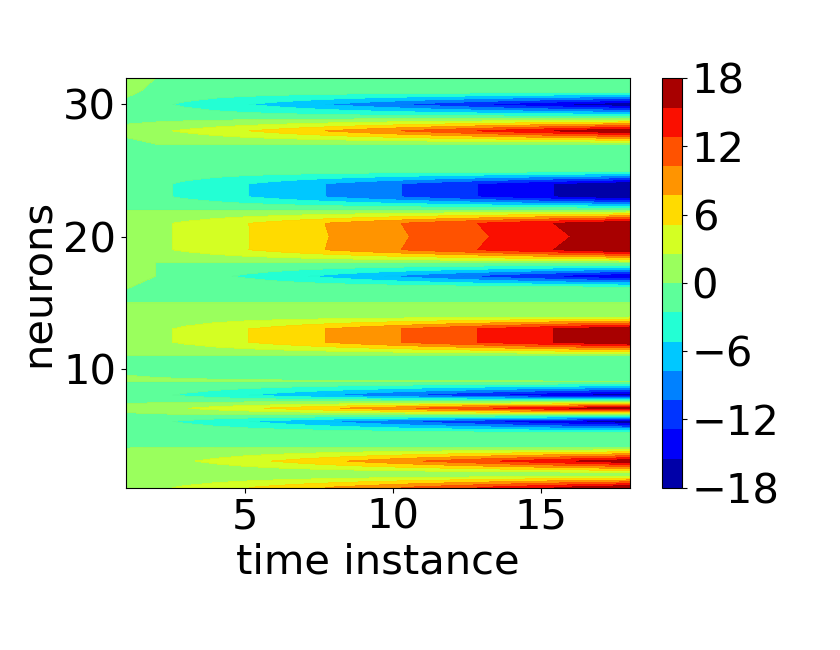}
    }
        \caption{Neuron activations of each gate $\mathbf{f}_{uv}(i), \mathbf{g}_{uv}(i), \mathbf{h}_{uv}(i), \mathbf{i}_{uv}(i), \mathbf{o}_{uv}(i)$, and $\mathbf{z}_{uv}(i)$ over time of the LSTM layer inside the CRNN model for two particular links $(u, v)$ in $algo$. The fact that several gate dimensions are non-zero indicates that information is propagating across multiple time periods for prediction. The top row demonstrates activations for a link formed late in the course, and the bottom row demonstrates activations for an early-formed link.}
        \label{neurons}
        \vspace{-0.15in}
\end{figure*}


\section{Link Formation Analytics}
\label{sec:recommending}

In this section, we consider several descriptive analytic tools and visualizations for instructors. We first describe the evolution of model parameters during prediction (Sec. \ref{sec:TSE}). We then examine the correlations between features (Sec. \ref{sec:featurecorrelations}) and analyze their individual and collective impact on prediction (Sec. \ref{sec:featureimportanceanalysis}). Finally, we analyze the importance of the predictor's architecture in Sec. \ref{sec:moel_arch}.

\vspace{-0.2cm}

\subsection{Time-Series Variable Evolution}
\label{sec:TSE}
Because the hidden layers of deep-learning models cannot be understood intuitively, we provide an alternate form of visualizing their behavior. It is possible to observe the decisions made by the deep learning model during prediction by investigating changes in state for each model gate over time, and making inferences about the final prediction from these observations. The stability exhibited by the gates over time supports the viability of early link formation prediction from Sec. \ref{sec:TAC}. To demonstrate this, we consider an example of how the $\mathtt{CRNN}$ LSTM layer parameters specified in Sec. \ref{sec:linkpredictionmethodology} for deep learning prediction models evolve over time. 

By examining the relationship fading gate, $\mathbf{f}$, in particular, we are able to demonstrate how the inputs from time interval $i -1$ affect the model output at time interval $i$, i.e., how much information is carried over from interval to interval. To do so, we choose a link $(u, v) \in \mathcal{G}(L)$ at random from $\mathtt{algo}$, and feed $\mathbf{e}_{uv}(i)$ into the trained model for $L = 20$ to generate the predictions $\tilde{y}_{uv}(i)$. The prediction has high accuracy on the chosen link, which forms within one time interval of when it is predicted to form.

The neuron activation values for the gates $\mathbf{g}$, $\mathbf{i}$, $\mathbf{f}$, $\mathbf{o}$ and the state $\mathbf{z}$ and output $\mathbf{h}$ are additionally considered and shown in Fig. \ref{neurons}. The vertical axis is the vector dimension (i.e., neuron number), and the horizontal is the time instance $i$. A few of the input gate dimensions, $\mathbf{g}$, change at about the time the link is formed (around $i = 17$). These changes propagate through the network, causing the output, $\mathbf{h}$, as well as some dimensions of the intermediate gates (e.g., $\mathbf{f}$, $\mathbf{i}$, and $\mathbf{o}$) to change around $i = 17$ as well, thus forming an accurate prediction. The fact that $\mathbf{i}$ and $\mathbf{f}$ in particular tend to take extreme values indicates that the input, $\mathbf{g}$, and prior state, $\mathbf{z}$, are either fully passed or blocked.

We also observe that several dimensions in $\mathbf{z}$ evolve gradually over time, with several non-zero dimensions in $\mathbf{f}$ passing information across multiple time periods. This result helps explain why models using an LSTM layer in conjunction with other methods perform better than the Bayesian model: passing information from one time interval to another increases the prediction quality compared to only updating the input features at each time interval.

\vspace{-0.2cm}

\subsection{Feature Correlations}
\label{sec:featurecorrelations} 
Investigating the relationship between individual features provides insights into the shape of an SLN in a different capacity than the predictions made by our deep-learning models, and it provides an analytical tool with which instructors can monitor an online classroom. Table \ref{summary_stats} summarizes the distributions of $\mathcal{G}(L)$ (top row) and $\mathcal{G}^{c}(L)$ (bottom row), with the top 5\% of outliers removed. We show the means and standard deviations (s.d.) of each feature for both groups, as well as the signal-to-noise ratio (SNR) for each feature. The large difference in magnitude for both mean and s.d. between formed and unformed links indicates a clear difference in behavior between these two groups. The large gap in values reinforces the results of our predictive algorithms discussed in Sec. \ref{sec:modelevaluationproc}. The SNR measures how effectively a feature can distinguish between the two groups, with a higher magnitude indicating more efficacy \cite{SNR}. We make a few impactful observations for link prediction from these statistics: 

\textit{(i) Infrequent short paths}: The length and number of shortest paths between learners are both negatively associated with link formation. The former is consistent with the intuition that learners who are closer together (i.e., smaller shortest path lengths) are more likely to form links. The latter, however, indicates that links are more likely to form when fewer such shortest paths exist, i.e., the paths should be unique. An interesting analogy can be drawn here to the small world phenomenon, where users can discover short paths in a social network even when only one or a few exist \cite{NPLP}; in other words, the presence of fewer short paths makes each of those neighboring connections more important and more likely to foster link creation.

\textit{(ii) Low-degreed shared neighbors}: In order of increasing SNR, $\mathtt{Ja}$, $\mathtt{Re}$ and $\mathtt{Ad}$ are each positively associated with link formation. Each of these measures the common neighborhood of two learners, with increasing penalty placed on the degrees of these neighbors (i.e., $\mathtt{Ja}$ does not include degree at all, while $\mathtt{Re}$ is inversely proportional to it). The fact that $\mathtt{Ad}$ has the highest SNR, then, implies that shared neighbors with fewer links are more prone to facilitate link formation, which is consistent with the the point above on unique paths being more predictive.

\textit{(iii) Low ceiling feature values}: Taking the statistics present in Table \ref{summary_stats} in conjunction with each feature's cumulative distribution function (CDF), shown in Fig. \ref{cdfs}, it is evident for several features including $\mathtt{To}$ and $\mathtt{Pr}$ that no learner pairs reach the maximum possible value for the feature. Most notably with respect to $\mathtt{To}$, the maximum number of shared topics between two connected users is always less than 15 of the 20 extracted topics. Given the highly connected nature of ``hub" students that possess a large number of shortest path connections, it would be expected that the maximum number of shared topics would be 20. This discrepancy in number of shared topics suggests that hub students connect frequently with less-engaged students, but rarely interact with each other, creating smaller student ecosystems within the course centered around their knowledge dissemination. Another possibility is a difference in student knowledge state/engagement on particular topics, indicating that learners are more motivated to post about topics they are confident in or interested in learning and avoid topics they are not.

\textit{(iv) Topology vs. post properties}: $\mathtt{Pr}$ and $\mathtt{To}$ are both positively associated with link formation, as one would expect: those with higher degrees ($\mathtt{Pr}$) and focusing on similar topics ($\mathtt{To}$) should be more likely to interact in the discussions. Surprisingly, though, these features have lower SNRs than the other neighborhood-based features, indicating that the network topology drives link formation in an SLN more than individual learner properties like a learner's tendency to post, for example, or topic interest. Furthermore, the SNR of $\mathtt{To}$ is higher in the less densely populated courses (\texttt{f19} and \texttt{s20}), indicating that clearer signals may emerge around topics when there is less overall volume of discussion in the forums. This is consistent with the performance differential of the GNN model in link prediction on the large vs. small datasets, since it does not learn from topic features.

\textit{(v) Quantitative vs. humanities courses}: Among the four MOOC courses, $\mathtt{Pr}$ is higher in $\mathtt{comp}$ and $\mathtt{shake}$ (particularly $\mathtt{shake}$) than in $\mathtt{ml}$ and $\mathtt{algo}$. This is consistent with humanities courses tending to invite more open-ended discussions, whereas quantitative courses have questions requiring explicit answers \cite{moocopt}. More learners would then be motivated to post in the forums of humanities courses -- in fact, such participation may be a course requirement -- leading to more links forming. Table \ref{course_stats} confirms the intuition that even with a smaller class size, $\mathtt{comp}$ and $\mathtt{shake}$ have a higher ratio of learner pairs to learners. The distinction between quantitative and humanities courses also helps explain which settings temporal behavior is helpful for link prediction, as we will discuss in Sec. \ref{sec:moel_arch}.

\subsection{Feature Importance Analysis}
\label{sec:featureimportanceanalysis}

\begin{table*}[!ht]
\centering
\begin{tabular}{c c c c c c c c} 
 \hline
 \multicolumn{2}{c}{Set} & $\mathtt{ml}$ & $\mathtt{algo}$ & $\mathtt{shake}$ & $\mathtt{comp}$ & $\mathtt{f19}$ & $\mathtt{s20}$\\
  \hline
 \hline
 \multirow{2}{1.5cm}{$\mathtt{Nei + Path}$} & AUC & \textbf{0.9487 $\pm$ 0.0241} & \textbf{0.9647 $\pm$ 0.0091} & 0.8978 $\pm$ 0.0303 & \textbf{0.9609 $\pm$ 0.0093} & \textbf{0.8945 $\pm$ 0.0330} & \textbf{0.9035 $\pm$ 0.0261} \\ 
  & ACC & \textbf{0.9528 $\pm$ 0.0196} & \textbf{0.9844 $\pm$ 0.0035} & \textbf{0.9693 $\pm$ 0.0071} & \textbf{0.9801 $\pm$ 0.0044} & \textbf{0.9695 $\pm$ 0.0064} & \textbf{0.9732 $\pm$ 0.0027} \\
  \hline
 \multirow{2}{1.5cm}{$\mathtt{Nei + Post}$} & AUC & 0.9398 $\pm$ 0.0011 & 0.9399 $\pm$ 0.0015 & 0.8541 $\pm$ 0.0024 & 0.8922 $\pm$ 0.0078 & 0.6735 $\pm$ 0.0519 & 0.6346 $\pm$ 0.0118 \\ 
  & ACC & 0.9446 $\pm$ 0.0008 & 0.9753 $\pm$ 0.0006 & 0.9314 $\pm$ 0.0050 & 0.9482 $\pm$ 0.0029 & 0.9538 $\pm$ 0.0015 & 0.9627 $\pm$ 0.0011  \\
  \hline
   \multirow{2}{1.5cm}{$\mathtt{Path + Post}$} & AUC & 0.9332 $\pm$ 0.0034 & 0.9455 $\pm$ 0.0058 & \textbf{0.9255 $\pm$ 0.0096} & 0.9444 $\pm$ 0.0078 & \textbf{0.8832 $\pm$ 0.0358} & 0.8848 $\pm$ 0.0175 \\
  & ACC & 0.9418 $\pm$ 0.0031 & 0.9659 $\pm$ 0.0028 & \textbf{0.9650 $\pm$ 0.0038} & 0.9736 $\pm$ 0.0039 & \textbf{0.9679 $\pm$ 0.0051} & \textbf{ 0.9736 $\pm$ 0.0022}\\
 \hline
\end{tabular}
\caption{Performance of the $\mathtt{CRNN}$ Model with selected input feature groups. The top two highest performing groups for each course metric are bolded. The combinations of $\mathtt{Nei + Path}$ and $\mathtt{Path + Post}$ outperform $\mathtt{Nei + Post}$ consistently, indicating that while neighborhood-based features are most important for prediction, the other feature types contribute significantly to link prediction as well.}
\label{feat_import_analysis}
\end{table*}

Recall in Sec. \ref{sec:featureengineering} that we define three groups of features: (i) $\mathtt{Nei}$, which quantify the overlap between learner neighborhoods, (ii) $\mathtt{Path}$, which are the length and number of shortest paths, and (iii) $\mathtt{Post}$, or the similarity in what learners discuss. To complement the correlation analysis in Table \ref{summary_stats} that was done for each feature individually, we now analyze the contribution of each feature type to the  prediction quality of our CRNN model, by evaluating it using different input feature combinations. 

To evaluate smaller groups of features using our CNN and CRNN models, a modification in model architecture is required. Our implementation of the CRNN model for computing links with all features contained both a $3 \times 1$ kernel layer and a $2 \times 1$ kernel layer. To classify samples using a subset of less than five of the seven features, the second convolutional layer using a $2 \times 1$ kernel was removed, leaving a single convolutional layer with a $3 \times 1$ kernel before the fully connected and output layers. This eliminates the issue of convolving a $1 \times 1$ output shape with an additional $2 \times 1$ kernel without requiring zero-padding. Determining the individual and combined effects of each feature group allows identification of potentially redundant features, which can improve computational speed when updating predictions in real time.

Table \ref{feat_import_analysis} shows the results when each course is broken into 20 time periods. None of the combinations reach the performance of the original model with all input variables in Table IV, indicating that each feature group contributes to the prediction quality. The $\mathtt{Nei + Path}$ and $\mathtt{Path + Post}$ combinations show the highest overall performance across all six forums, indicating that the combination of $\mathtt{Nei + Path}$ has a confounding effect on the model -- we would expect both $\mathtt{Nei}$-based groups to share a higher AUC. Combining these values with the SNRs in Table \ref{summary_stats} indicates that the $\mathtt{Nei}$ features contribute the most to model accuracy, followed by $\mathtt{Post}$ and then $\mathtt{Path}$. 

If we compare the individual feature groups, we generally find that the $\mathtt{Nei}$ features perform the best, followed by $\mathtt{Path}$, and then $\mathtt{Post}$. This is consistent with the behavior of these features within groups as well. This ordering of $\mathtt{Post}$ and $\mathtt{Path}$ is opposite of the SNR magnitudes from Table \ref{summary_stats}: here, the single feature $\mathtt{To}$ outperforms the combined impact of $\mathtt{Path}$. Given that Table \ref{summary_stats} is concerned with the eventual formation of links but not the time at which they form, we conjecture that in the absence of $\mathtt{Nei}$, $\mathtt{Post}$ is more important to pinpointing the time of link formation while $\mathtt{Path}$ is more important to whether they form at all. After all, the timing of particular topic coverage should influence when learners interested in those topics connect.

\vspace{-0.2cm}

\subsection{Model Architecture Analysis}
\label{sec:moel_arch}

Here, we first analyze the importance of spatial pattern preserving convolutional layers and temporal pattern preserving recurrent layers for link prediction in SLNs. We find that, in general, classification models that incorporate only spatial pattern dependencies ($\mathtt{CNN}$) outperform models that only incorporate time dependencies ($\mathtt{RNN}$), as shown in Table \ref{model_perf}. This is consistent with Table \ref{feat_import_analysis}, where we find that SLN topology features (i.e., neighborhood and path-based features), which explain spatial relationships between links, are the most important for accurate link prediction. However, we also find that incorporating time dependencies into link prediction models (e,g., $\mathtt{RNN}$ and $\mathtt{CRNN}$) obtains strong performance in large courses such as MOOCs, whereas these models become less accurate on small courses such as $\mathtt{f19}$ and $\mathtt{s20}$. Interestingly, although the $\mathtt{RNN}$ accurately predicts whether links will form (as shown in Table \ref{model_perf}), they do not accurately predict when the links will form as shown from the TAC curves in Fig. \ref{tac}, particularly on $\mathtt{ml}$, $\mathtt{f19}$, and $\mathtt{s20}$. This behavior is consistent with such quantitative courses requiring short answers in fast time intervals whereas the humanities courses typically involve threads of discussion that persist over longer periods of time \cite{sln_eff}. In Fig. \ref{neurons}, we further explored the efficacy of recurrent layers by visualizing the various gates of the $\mathtt{CRNN}$ in the $\mathtt{algo}$ course, where we saw that information propagates from multiple time periods to aid link prediction after spatial patterns have been identified. This reinforces that recurrent layers may carry long-term information for link prediction, but convolutional layers are more robust for in SLNs on both large and small courses.

In addition, as shown in Table \ref{model_perf}, convolutional GNNs achieve strong link prediction performance on each of the MOOC datasets. Rather than employing our explicitly defined model features, GraphSAGE embeds features across the SLN topology that exploit spatial patterns, hence resulting in strong performance for these datasets captured by the GNN's convolutional layers. However, for the GNN to learn such discriminative features, it may require a large graph to train on \cite{gnns}, thus making the model less effective for smaller courses such as $\mathtt{f19}$ and $\mathtt{s20}$. Our proposed framework, in which we explicitly model features between node pairs, on the other hand, is better able to learn and generalize on the smaller datasets. More generally, these results indicate that in the SLN domain, informed feature engineering (i.e., using spatial features) paired with corresponding layers (i.e., convolutional) results in better trained models with less data than that required by GNNs. This is useful for generating analytics in the early stages of courses before a significant amount of links have formed (i.e., before interaction data has been observed) on the forums \cite{moocperformance,moocopt}.

\vspace{-0.3cm}

\section{Conclusion}
\label{sec:conclusion}

In this work, we developed a link prediction framework specifically tailored to operate in social learning networks (SLNs) based on neighborhood-based, path-based, and post-based modeling features. Through evaluation in six different courses, we demonstrated our framework's ability to perform accurate link prediction in a variety of learning environments. In particular, we examined the efficacy of our framework on a course forced online after approximately eight weeks of traditional instruction due to the COVID-19 pandemic. In addition, we considered the SLNs formed in four Massive Open Online Courses (MOOCs) as well as one traditional undergraduate course, with a heavy reliance on student participation in an online discussion forum, offered through Purdue University.

While our work establishes an initial framework and results for link prediction in SLNs, many avenues remain for exploring the challenges of link prediction in this new type of online social network. One is additional feature engineering: other features that we did not consider -- such as learners’ background knowledge, level of education, and personal goals -- may also be associated with link formation, and may allow further improvements in link prediction quality. As demonstrated here, our proposed framework is applicable across multiple datasets; thus, additional evaluation variants on forums or classes with different structures, such as those present in K-12 education, may be beneficial.


\end{document}